\documentclass[%
a4paper,							% paper format
11pt,								% Schriftgröße (12pt, 11pt (Standard))
bibliography=totoc,						% Literaturverzeichnis im Inhalt
abstracton,					% Überschrift über der Zusammenfassung an	
]
{scrartcl}

\usepackage[a4paper,left=3.0cm,right=2.5cm, top=2.5cm, bottom=3.0cm]{geometry}
\usepackage[headsepline=.4pt,footsepline=.4pt,automark,autooneside=false,]{scrlayer-scrpage}
\clearscrheadfoot 
\automark[subsection]{section} 
\automark*[subsubsection]{subsection}
\ihead{\scriptsize\rightmark} 
\usepackage[ngerman, english]{babel}
\usepackage[T1]{fontenc}
\usepackage[utf8]{inputenc}
\usepackage{lmodern} %latin modern
\usepackage[onehalfspacing]{setspace} %1,5 spacing

\usepackage{xspace}
\usepackage{calc}%\widthof{}

\usepackage{amsmath}
\usepackage{amssymb}
\usepackage{amsthm,thmtools}

\usepackage{mathtools}
\usepackage{bbm}
\begingroup
    \makeatletter
    \@for\theoremstyle:=definition,remark,plain\do{%
        \expandafter\g@addto@macro\csname th@\theoremstyle\endcsname{%
            \addtolength\thm@preskip\parskip
            }%
        }
\endgroup

\usepackage{chngcntr}%reset equation counter after each section
\counterwithin*{equation}{section}

\theoremstyle{plain}
\newtheorem{theorem}{Theorem}[section]
\newtheorem{lemma}[theorem]{Lemma}
\theoremstyle{definition}
\newtheorem{definition}[theorem]{Definition}

\theoremstyle{remark}
\newtheorem{remark}[theorem]{Remark}
\theoremstyle{plain}
\newtheorem{proposition}[theorem]{Proposition}
\theoremstyle{plain}

\theoremstyle{remark}

\usepackage{tabularx}
\usepackage{booktabs}
\usepackage{multirow}
\usepackage{multicol}
\usepackage{pdflscape}

\usepackage{diagbox}
\usepackage[flushleft]{paralist}
\setdefaultenum{(i)}{(a)}{(1)}{(aa)}%default numbering in law

\usepackage{color}
\usepackage{graphicx}
\usepackage{tikz}
\usetikzlibrary{calc,spy,backgrounds}
\usepackage[final]{showkeys}

\usepackage{fancyvrb}
\usepackage{float}
\usepackage[section]{placeins}
\usepackage[format=plain]{caption}
\usepackage{titleref}
\usepackage{hyperref}
\usepackage{cleveref}
\hypersetup{
    colorlinks,
		linkcolor={red},
    citecolor={blue},
    urlcolor={blue}
}
\renewcommand{\tilde}[1]{\widetilde{#1}}
\newcommand{\PCIR}{P^\text{CIR-}}
\newcommand{\rCIR}{r^\text{CIR-}}

\newcommand{\Swap}{\mathrm{Swap}}
\newcommand{\parSwapRate}{\mathrm{R}}
\newcommand{\PVBP}{\mathrm{S}}

\newcommand{\Swaption}{\mathrm{Swaption}}

\newcommand{\BSwaption}{\mathrm{BSwaption}}
\newcommand{\swapType}{\zeta}%+1 payer, -1 receiver
\newcommand{\abrBSwaption}[2]{$#1\,\mathrm{nc}\,#2$}

\newcommand{\CMS}{\mathrm{CMS}}
\newcommand{\R}{\mathbb{R}}
\newcommand{\Q}{\mathbb{Q}}
\newcommand{\N}{\mathbb{N}}
\newcommand{\1}{\mathbbm{1}}
\newcommand{\normalCDF}{\mathcal{N}}
\newcommand{\normalPDF}{\varphi}

\newcommand{\norm}[1]{\left\|#1\right\|}
\newcommand{\abs}[1]{\left|#1\right|}
\DeclareMathOperator*{\argmin}{arg\,min}

\newcommand{\tabEC}[2]{$f(\Pi)=$ #1\newline in #2\,s}% for calibration table
\newcolumntype{C}{>{\centering\arraybackslash}X}

\newcommand{\CPU}{Intel(R) Core(TM) i7-8750H CPU @ 2.20\,GHz\xspace}
\newcommand{\RAM}{2x32\,GB (Dual Channel) Samsung SODIMM DDR4 RAM @ 2667 MHz\xspace}

\newcommand{\OS}{Windows 10 Pro\xspace}

\SaveVerb{matlab}=Matlab 2021a=
\newcommand{\matlab}{\protect\UseVerb{matlab}\xspace}
\SaveVerb{ga}=ga=
\newcommand{\ga}{\protect\UseVerb{ga}\xspace}
\SaveVerb{fmincon}=fmincon=
\newcommand{\fmincon}{\protect\UseVerb{fmincon}\xspace}
\newcommand{\matlabGOtoolbox}{(Global) Optimization Toolbox\xspace}
\newcommand{\dateA}{30/12/2019\xspace}
\newcommand{\dateB}{30/11/2020\xspace}
\newcommand{\marketDataA}{
\begin{tabular}{|*{3}{c}|}
\hline
Maturity (in years) & Zero rate (in \%) & Zero-coupon price\\
\hline
$0.0833333333333333$ & $-0.469999993219972$ & $  1.0004001991529$ \\
$             0.25$ & $-0.388000020757318$ & $ 1.00096969387991$ \\
$              0.5$ & $-0.324999983422458$ & $ 1.00163343819125$ \\
$             0.75$ & $-0.314333918504417$ & $ 1.00237481461989$ \\
$                1$ & $-0.322000007145107$ & $ 1.00323926670136$ \\
$             1.25$ & $-0.323286440253412$ & $ 1.00405360258242$ \\
$              1.5$ & $-0.316161320131414$ & $ 1.00476558980205$ \\
$             1.75$ & $-0.303842297803669$ & $ 1.00535001652119$ \\
$                2$ & $-0.289547047577798$ & $ 1.00582418019158$ \\
$             2.25$ & $-0.275860329135469$ & $ 1.00623288634409$ \\
$              2.5$ & $-0.262835313503729$ & $   1.006604855007$ \\
$             2.75$ & $-0.249892233800608$ & $ 1.00691299093433$ \\
$                3$ & $-0.236451346427202$ & $ 1.00713375064174$ \\
$             3.25$ & $-0.222084053437044$ & $ 1.00725039326453$ \\
$              3.5$ & $-0.20696636298112$ & $ 1.00728054250496$ \\
$             3.75$ & $-0.191425434683623$ & $ 1.00721781901104$ \\
$                4$ & $-0.175788428168744$ & $ 1.00706740209126$ \\
$             4.25$ & $-0.160311330630236$ & $ 1.00684531811395$ \\
$              4.5$ & $-0.144965462482105$ & $ 1.00655553463348$ \\
$             4.75$ & $-0.129650957156002$ & $ 1.00618948972951$ \\
$                5$ & $-0.114267959725112$ & $ 1.00573933685071$ \\
$             5.25$ & $-0.0987154224631581$ & $ 1.00520062530541$ \\
$              5.5$ & $-0.0828875612342017$ & $ 1.00457454544122$ \\
$             5.75$ & $-0.0666773874613114$ & $ 1.00384671986489$ \\
$                6$ & $-0.0499779242090881$ & $ 1.00300667524933$ \\
$             6.25$ & $-0.0327643402378897$ & $ 1.00205088034181$ \\
$              6.5$ & $-0.0153403983915723$ & $ 1.00099833086134$ \\
$             6.75$ & $0.00190798987986796$ & $0.999871102605028$ \\
$                7$ & $0.0185949131264351$ & $0.998698306220564$ \\
$             7.25$ & $0.0344518735623467$ & $0.997505079039002$ \\
$              7.5$ & $0.0496800311054812$ & $0.996279818846146$ \\
$             7.75$ & $0.0645979575189415$ & $0.995003816465917$ \\
$                8$ & $0.0795242260210216$ & $0.993656440330286$ \\
$             8.25$ & $0.0947347900819295$ & $0.992214008696662$ \\
$              8.5$ & $0.110335148849572$ & $0.990662992494919$ \\
$             8.75$ & $0.126388167535652$ & $0.988997743889118$ \\
$                9$ & $0.142956722993404$ & $0.987213788328959$ \\
$             9.25$ & $0.160050573928316$ & $0.985308478446392$ \\
$              9.5$ & $0.177466994199449$ & $0.983284710270437$ \\
$             9.75$ & $0.194950156980411$ & $0.981173005874126$ \\
$               10$ & $0.212244223803282$ & $0.979004189945635$ \\
$               15$ & $0.473523046821356$ & $0.931543316237289$ \\
$               20$ & $0.611338950693607$ & $0.885166902653398$ \\
$               25$ & $0.652327481657267$ & $0.849865688031976$ \\
$               30$ & $0.640345783904195$ & $0.825611308910539$ \\\hline
\end{tabular}
}

\newcommand{\marketSwaptionA}{
\begin{tabular}{|c|*{5}{c}|}
\hline
\diagbox{Maturity}{Tenor} & 1 & 2 & 5 & 7 & 10 \\
 \hline
$1$ & $0.000702236$ & $0.00175071$ & $0.00706456$ & $0.0112631$ & $0.0181169$ \\
$2$ & $0.0014433$ & $0.00333027$ & $0.0111956$ & $0.017189$ & $0.0265694$ \\
$5$ & $0.00391314$ & $0.00796766$ & $0.0214221$ & $0.0308074$ & $0.0453508$ \\
$7$ & $0.00521117$ & $0.0104082$ & $0.0268942$ & $0.0380283$ & $0.0548627$ \\
$10$ & $0.00668368$ & $0.0132567$ & $0.0330802$ & $0.045932$ & $0.0651091$ \\
$15$ & $0.00781681$ & $0.0154396$ & $0.0378811$ & $0.0525334$ & $0.0743464$ \\
$20$ & $0.00840243$ & $0.0166069$ & $0.0407885$ & $0.0565876$ & $0.0795953$ 
\\\hline
\end{tabular}
}

\newcommand{\strikeSwaptionA}{
\begin{tabular}{|c|*{5}{c}|}
\hline
\diagbox{Maturity}{Tenor} & 1 & 2 & 5 & 7 & 10 \\
 \hline
$1$ & $-0.260793\,\% $ & $-0.195187\,\% $ & $-0.011405\,\% $ & $0.140129\,\% $ & $0.330514\,\% $ \\
$2$ & $-0.129665\,\% $ & $-0.0782444\,\% $ & $0.139932\,\% $ & $0.273273\,\% $ & $0.449172\,\% $ \\
$5$ & $0.268095\,\% $ & $0.38307\,\% $ & $0.556996\,\% $ & $0.655339\,\% $ & $0.757978\,\% $ \\
$7$ & $0.547079\,\% $ & $0.611571\,\% $ & $0.76683\,\% $ & $0.830788\,\% $ & $0.891069\,\% $ \\
$10$ & $0.880582\,\% $ & $0.907944\,\% $ & $0.967521\,\% $ & $0.988131\,\% $ & $0.992003\,\% $ \\
$15$ & $1.04232\,\% $ & $1.04153\,\% $ & $1.01776\,\% $ & $0.985317\,\% $ & $0.924744\,\% $ \\
$20$ & $0.925377\,\% $ & $0.901441\,\% $ & $0.827386\,\% $ & $0.778437\,\% $ & $0.721445\,\% $ 
\\\hline
\end{tabular}
}

\newcommand{\volSwaptionA}{
\begin{tabular}{|c|*{7}{c}|}
\hline
\diagbox{Maturity}{Tenor} & 1 & 2 & 3 & 4 & 5 & 7 & 10 \\
 \hline
$1$ & $17.5$ & $21.8$ & $26.8$ & $31.4$ & $35.2$ & $40.2$ & $45.6$ \\
$2$ & $25.4$ & $29.3$ & $33.5$ & $36.4$ & $39.5$ & $43.5$ & $47.5$ \\
$3$ & $34$ & $36.7$ & $39.2$ & $41.1$ & $43.2$ & $46.2$ & $49.3$ \\
$4$ & $40$ & $41.5$ & $43.4$ & $44.8$ & $46.2$ & $48.4$ & $50.9$ \\
$5$ & $43.7$ & $44.6$ & $45.8$ & $47$ & $48.4$ & $50.1$ & $52.3$ \\
$7$ & $49.7$ & $49.8$ & $50.5$ & $51.4$ & $52.1$ & $53.1$ & $54.4$ \\
$10$ & $54.6$ & $54.4$ & $54.7$ & $54.9$ & $55.1$ & $55.2$ & $55.6$ \\
$15$ & $54.8$ & $54.4$ & $54.5$ & $54.4$ & $54.2$ & $54.2$ & $54.4$ \\
$20$ & $53.6$ & $53.2$ & $53.4$ & $53$ & $52.9$ & $52.8$ & $52.5$ 
\\\hline
\end{tabular}
}

\newcommand{\bSwaptionStrikesA}{
\begin{tabular}{|c|*{4}{c}|}
\hline
\diagbox{Maturity}{Tenor} & 2 & 5 & 7 & 10\\
\hline
1 & -0.194\,\% & 0.00912\,\% & 0.14\,\% & 0.33\,\%\\
3 & 0.0789\,\% & 0.274\,\% & 0.432\,\% & 0.561\,\%\\
5 & 0.335\,\% & 0.534\,\% & 0.644\,\% & 0.767\,\%\\
7 & 0.612\,\% & 0.771\,\% & 0.84\,\% & 0.894\,\%\\
10 & 0.926\,\% & 1.01\,\% & 0.994\,\% & 1.02\,\%\\\hline
\end{tabular}
}
\newcommand{\receiverbSwaptionHWA}{
\begin{tabular}{|c|*{4}{c}|}
\hline
\diagbox{Maturity}{Tenor} & 2 & 5 & 7 & 10\\
\hline
1 & 0.21\,\% & 1.06\,\% & 1.85\,\% & 3.28\,\%\\
3 & 0.57\,\% & 1.83\,\% & 2.86\,\% & 4.63\,\%\\
5 & 0.87\,\% & 2.48\,\% & 3.71\,\% & 5.72\,\%\\
7 & 1.11\,\% & 3.03\,\% & 4.43\,\% & 6.65\,\%\\
10 & 1.4\,\% & 3.62\,\% & 5.2\,\% & 7.59\,\%\\\hline
\end{tabular}
}
\newcommand{\payerbSwaptionHWA}{
\renewcommand{\arraystretch}{1}
\begin{tabular}{|c|*{4}{c}|}
\hline
\diagbox{Maturity}{Tenor} & 2 & 5 & 7 & 10\\
\hline
1 & 0.25\,\% & 1.4\,\% & 2.55\,\% & 4.76\,\%\\
3 & 0.6\,\% & 2.08\,\% & 3.42\,\% & 5.74\,\%\\
5 & 0.9\,\% & 2.7\,\% & 4.16\,\% & 6.59\,\%\\
7 & 1.13\,\% & 3.2\,\% & 4.75\,\% & 7.18\,\%\\
10 & 1.41\,\% & 3.72\,\% & 5.33\,\% & 7.91\,\%\\\hline
\end{tabular}
}

\title{On the deterministic-shift extended CIR model in a negative interest rate framework}
\author{Marco Di Francesco\thanks{UnipolSai Assicurazioni, via Stalingrado 45, Bologna, Italy, \textbf{e-mail}: marco.difrancesco@unipolsai.com}\and Kevin Kamm\thanks{Dipartimento di Matematica, Universit\`a di Bologna, Bologna, Italy.
\textbf{e-mail}: kevin.kamm@unibo.it}}

\begin{document}
%% roman numbering
%%%%%%%%%%%%%%%%%%%%%%%%%%%%%%%%%%%%%%%%%%%%%%%%%%%%%%%%%%%%%%%%%%%%%%%%%%%
\thispagestyle{empty}\pagenumbering{roman}
%%%%%%%%%%%%%%%%%%%%%%%%%%%%%%%%%%%%%%%%%%%%%%%%%%%%%%%%%%%%%%%%%%%%%%%%%%%%%%%%%%%%%%%%%%%
%% Titlepage
%%%%%%%%%%%%%%%%%%%%%%%%%%%%%%%%%%%%%%%%%%%%%%%%%%%%%%%%%%%%%%%%%%%%%%%%%%%%%%%%%%%%%%%%%%%
\maketitle
\renewcommand{\thefootnote}{\Roman{footnote}}
\footnotetext[0]{The views expressed in this note are the only responsibility of the author and do
not represent in any way those of author's current employer. All errors are the only
responsibility of the author.}
\renewcommand{\thefootnote}{\arabic{footnote}}
%\newpage
%%%%%%%%%%%%%%%%%%%%%%%%%%%%%%%%%%%%%%%%%%%%%%%%%%%%%%%%%%%%%%%%%%%%%%%%%%%%%%%%%%%%%%%%%%%
%% Abstract
%%%%%%%%%%%%%%%%%%%%%%%%%%%%%%%%%%%%%%%%%%%%%%%%%%%%%%%%%%%%%%%%%%%%%%%%%%%%%%%%%%%%%%%%%%%
\begin{abstract}
In this paper, we propose a new exogenous model to address the problem of negative interest rates that preserves the analytical tractability of the original Cox-Ingersoll-Ross (CIR) model with a perfect fit to the observed term-structure. We use the difference of two independent CIR processes and apply the deterministic-shift extension technique.
To allow for a fast calibration to the market swaption surface, we apply the Gram-Charlier expansion to calculate the swaption prices in our model. We run several numerical tests to demonstrate the strengths of this model by using Monte-Carlo techniques. In particular, the model produces close Bermudan swaption prices compared to Bloomberg's Hull-White one-factor model. Moreover, it finds constant maturity swap (CMS) rates very close to Bloomberg's CMS rates. 
\end{abstract}
%%%%%%%%%%%%%%%%%%%%%%%%%%%%%%%%%%%%%%%%%%%%%%%%%%%%%%%%%%%%%%%%%%%%%%%%%%%%%%%%%%%%%%%%%%%
%% Keywords
%%%%%%%%%%%%%%%%%%%%%%%%%%%%%%%%%%%%%%%%%%%%%%%%%%%%%%%%%%%%%%%%%%%%%%%%%%%%%%%%%%%%%%%%%%%
\noindent
\textbf{Acknowledgements:}
This project has received funding from the European Union’s Horizon 2020 research and innovation programme
under the Marie Sklodowska-Curie grant agreement No 813261 and is part of the ABC-EU-XVA
project.
\newpage
%%% Roman numbering
%%%%%%%%%%%%%%%%%%%%%%%%%%%%%%%%%%%%%%%%%%%%%%%%%%%%%%%%%%%%%%%%%%%%%%%%%%%%
%\thispagestyle{scrheadings}\ihead{}\pagenumbering{Roman}
%%%%%%%%%%%%%%%%%%%%%%%%%%%%%%%%%%%%%%%%%%%%%%%%%%%%%%%%%%%%%%%%%%%%%%%%%%%%%%%%%%%%%%%%%%%%
%%% Table of Contents
%%%%%%%%%%%%%%%%%%%%%%%%%%%%%%%%%%%%%%%%%%%%%%%%%%%%%%%%%%%%%%%%%%%%%%%%%%%%%%%%%%%%%%%%%%%%
%\tableofcontents
%\newpage
%% arabic numbering
%%%%%%%%%%%%%%%%%%%%%%%%%%%%%%%%%%%%%%%%%%%%%%%%%%%%%%%%%%%%%%%%%%%%%%%%%%%
\pagestyle{scrheadings}\ihead{\scriptsize\rightmark}\pagenumbering{arabic}
%%%%%%%%%%%%%%%%%%%%%%%%%%%%%%%%%%%%%%%%%%%%%%%%%%%%%%%%%%%%%%%%%%%%%%%%%%%%%%%%%%%%%%%%%%%
%% Arcticle Body
%%%%%%%%%%%%%%%%%%%%%%%%%%%%%%%%%%%%%%%%%%%%%%%%%%%%%%%%%%%%%%%%%%%%%%%%%%%%%%%%%%%%%%%%%%%
%%%%%%%%%%%%%%%%%%%%%%%%%%%%%%%%%%%%%%%%%%%%%%%%%%%%%%%%%%%%%%%%%%%%%%%%%%%%%%%%%%%%%%%%%%%
%% Introduction
%%%%%%%%%%%%%%%%%%%%%%%%%%%%%%%%%%%%%%%%%%%%%%%%%%%%%%%%%%%%%%%%%%%%%%%%%%%%%%%%%%%%%%%%%%%
\section{Introduction}\label{sec:introduction}
This paper is the natural extension of \cite{DiFrancesco2021} where we modeled 
interest rates by means of a short-rate model defined as the difference of two independent Cox-Ingersoll-Ross (CIR) processes in a negative interest rate framework.

We are extending the previous short-rate model by adding a deterministic function to allow for a perfect fit to the observed market term-structure while preserving its analytical tractability of an affine model and its features.

%%%%%%%%%%%%%%%%%%%%%%%%%%%%%%%%%%%%%%%%%%%%%%%%%%%%%%%%%%%%%%%%%%%%%%%%%%%%%%%%%%%%%%%%%%%
%% Little Description of the previous paper
Let us briefly recall our findings in \cite{DiFrancesco2021}. 
We derived an analytical formula for the zero-coupon price of the non-extended model (see \Cref{thm:zcpriceCIR-}) by solving the associated Riccati equations in \Cref{lem:Riccati} explicitly and calibrated it to the market term-structure.
Such short-rate models, where the observed term-structure is an output depending on the model parameters, are called \emph{endogeneous}.
We performed several numerical experiments at two different dates obtaining good results in the sense that the model reproduced the market term-structures with negative interest rates very well
and it generates more realistic distributions of interest rates with slight skewness and fatter tail with respect to the normal distribution. But, as reported in the numerical tests, the model failed to capture the full swaption surface due to the fact that the model parameters were constant and the Brownian motions were independent.

%In \cite{DiFrancesco2021} we derived the zero-coupon price of this model in \Cref{thm:zcpriceCIR-} by solving the associated Riccati equations in \Cref{lem:Riccati} explicitly and calibrated it to the market termstructure. The model fit the market termstructure well but lacked in matching the swaption surface.
%%%%%%%%%%%%%%%%%%%%%%%%%%%%%%%%%%%%%%%%%%%%%%%%%%%%%%%%%%%%%%%%%%%%%%%%%%%%%%%%%%%%%%%%%%%
%% Little Description of the paper
To improve the fit to the swaption surface, we suggest to transform the endogenous model into an exogenous one, in which the observed term-structure is an input. 

A basic strategy to transform an endogenous model to an exogenous one, is the inclusion of time-dependent parameters to exactly reproduce the observed term-structure. In fact, matching the term-structure exactly is equivalent to solving a system with an infinite number of equations. However, this is only possible by introducing an infinite number of parameters or, equivalently, a deterministic function of time. We follow the method illustrated in \cite[pp.~95\,ff. Section 3.8 A General Deterministic-Shift Extension]{BrigoMercurioLibro} to extend any time-homogeneous
short-rate model, so as to exactly reproduce any observed term-structure
of interest rates while preserving the possible analytical tractability of the original model.

%In this paper we will tackle this problem and use the deterministic shift extension technique by \cite{BrigoMercurioLibro} to guarantee a perfect fit to the market termstructure and calibrate the model to parts of the swaption surface.

To be more precise, we consider the CIR dynamics for $z\in \left\{x,y\right\}$
\begin{align}
 dz(t) = k_z(\theta_z - z(t))dt + \sigma_z \sqrt{z(t)} dW_z(t), \quad z(0)=z_0
\label{eq:CIR}
\end{align}
under a martingale measure $\Q$ with $k_z,\theta_z,\sigma_z \in \R_{>0}$ and define the short-rate as 
\begin{align}
	\label{eq:r}
	r(t)\coloneqq x(t)-y(t)+\psi\left(t\right),
\end{align}
where $W_y$ and $W_x$ are two independent standard Brownian motions on a stochastic basis 
$\left(\Omega,\mathcal{F},\left(\mathcal{F}_t\right)_{t\in [0,T]},\Q\right)$ and 
$\psi(t)\coloneqq f^M(0,t)-f(0,t)$ is a deterministic function defined as the difference of the market and model instantaneous forward rate. 

Since the market term-structure is now an input, we can calibrate the model parameters to the swaption surface. However, simple Monte-Carlo techniques are in general very slow and memory demanding. Therefore, we resort to an approximation formula known as Gram-Charlier expansion (cf. \cite{Tanaka2010}) in our model. This allows for a fast and accurate calibration procedure.

%Afterwards, we will use the calibrated model to determine constant maturity swap rates and price Bermudam swaptions to assess its performance.
%%%%%%%%%%%%%%%%%%%%%%%%%%%%%%%%%%%%%%%%%%%%%%%%%%%%%%%%%%%%%%%%%%%%%%%%%%%%%%%%%%%%%%%%%%%
%% Description of the main results
%%%%%%%%%%%%%%%%%%%%%%%%%%%%%%%%%%%%%%%%%%%%%%%%%%%%%%%%%%%%%%%%%%%%%%%%%%%%%%%%%%%%%%%%%%%
\subsection{Description of the main results}\label{sec:intro_description}
In this paper, we will first of all extend the results of \cite{DiFrancesco2021} by using a deterministic shift extension. The zero-coupon price in the extended model 
\eqref{eq:r} is given in the next Lemma.
\begin{lemma}\label{lem:zcPriceMain}%
		Let $\left(\Omega,\mathcal{F},\left(\mathcal{F}_t\right)_{t\in [0,T]},\Q\right)$ be a 
	stochastic basis, where $\Q$ is a martingale measure, $T>0$ a finite time horizon and let the $\sigma$-algebra $\left(\mathcal{F}_t\right)_{t\in [0,T]}$ fulfill the usual conditions and support two independent standard Brownian motions $W_x$ and $W_y$.
	The price of a zero-coupon bond in the model $r(t)\coloneqq x(t)-y(t)+\psi(t)$ is given by
	\begin{align*}
		P(t,T)=
		\frac{P^M(0,T)}{P^M(0,t)}\frac{\PCIR(0,t)}{\PCIR(0,T)}
		\PCIR(t,T),
	\end{align*}
	where $\PCIR(t,T)$ is the zero-coupon price from Theorem \ref{thm:zcpriceCIR-} and
	$P^M(0,T)$ the market zero-curve.
\end{lemma}
The derivation of this result is straightforward and referred to \Cref{sec:model} alongside a recollection of basic results on swaps and swaptions.

We will see that it is necessary to study the so-called \emph{swap moments} to derive the Gram-Charlier expansion. In our model, we will find explicit formulas allowing for fast swaption pricing and it is part of the next technical Lemma.
\begin{lemma}\label{lem:main}%
	Let everything be as in Lemma \ref{lem:zcPriceMain}. The so-called \emph{swap moments}
	at time $t<T_0$ of order $m\in \N$ are given by
	\begin{multline*}
		M^m(t)\coloneqq
	\mathbb{E}^{\Q^{T_0}}\left[
		\left.
		\left(
			\Swap_{T_0}^{T_N}(T_0;K,\swapType)
		\right)^m
		\right|
		\mathcal{F}_t
	\right]=
	\left(\frac{P^\text{CIR-}(0,T_0)}{P^M(0,T_0)}\right)^m
	\frac{1}{\PCIR\left(t,T_0\right)}\\
	\sum_{\substack{0\leq k_0,\dots,k_N\leq N\\k_0+\dots+k_N=m}}^{}{}
		\frac{m!}{k_0!\cdots k_N!}
		\tilde{a}_{0}^{k_0}\cdots \tilde{a}_{N}^{k_N} 
		\biggl(
			M_x(t,T_0) e^{-N_x(t,T_0) x(t)}M_y(t,T_0) e^{N_y(t,T_0) y(t)}
		\biggr)
	\end{multline*}
where we suppress the dependency of $N_z,M_z$ on $k_i$ for readability. The coefficients
$\tilde{a}_i$ are given by
\begin{align*}
	\tilde{a}_0&\coloneqq
		\swapType
		\frac{P^M(0,T_{0})}{P^\text{CIR-}(0,T_{0})},& 
	\tilde{a}_N&\coloneqq
		-\swapType\left(1+K\alpha_{N}\right)
		\frac{P^M(0,T_{N})}{P^\text{CIR-}(0,T_{N})},& 
	\tilde{a}_i&\coloneqq
		-\swapType K\alpha_i \frac{P^M(0,T_{i})}{P^\text{CIR-}(0,T_{i})},
\end{align*}
for $i=1,\dots,N-1$, year fractions $\alpha_i$, fixed swap rate $K$ and swap type $\swapType=1$ for a payer swap and $\swapType=-1$ for a receiver swap.

Moreover, the functions $M_z,N_z$, $z\in\left\{x,y\right\}$ are defined as
	\begin{align*}
	M_z(t,T_0)&=
		a_z
		\left(
			\frac{
				\phi_1^z \exp\left(\phi_2^z (T_0-t)\right)
			}{
				\phi_1^z+
				\phi_2^z \left(\exp\left(\phi_1^z (T_0-t)\right)-1\right)
					\left(1+b_z\left(\phi_1^z-\phi_2^z\right)\right)
			}
		\right)^{\phi_3^z},\
	a_z=
		\prod_{j=0}^{N}{
			A_z(T_0,T_{j})^{k_j} 
		}\\
	N_z(t,T_0)&=
		\frac{
			b_z \phi_1^z + 
			\left(\exp\left(\phi_1^z (T_0-t)\right)-1\right)
				\left(1+b_z\left(\phi_1^z-\phi_2^z\right)\right)
		}{
			\phi_1^z+
				\phi_2^z \left(\exp\left(\phi_1^z (T_0-t)\right)-1\right)
					\left(1+b_z\left(\phi_1^z-\phi_2^z\right)\right)
		},\
		b_z=
		\sum_{j=0}^{N}{k_j B_z(T_0,T_{j})},
	\end{align*}
	where $A_z,B_z$ are the functions defined in \Cref{thm:zcpriceCIR-}.
	The swap cumulants $c_l(t)$ at time $t$ are now given by the formulas in \Cref{sec:cumulants} by setting $\mu_i\coloneqq M^l(t)$, $l=1,\dots,m$.
\end{lemma}
For the proof of this Lemma we follow \cite{Tanaka2010} closely, which is referred to \Cref{sec:moments}.

The main result of this paper is the approximation of swaption prices by the Gram-Charlier expansion with short-rate \eqref{eq:r}, which follows immediately from 
\Cref{lem:main} by using \Cref{prop:GramCharlier} and is referred to \Cref{sec:expansionFormula}.
\begin{theorem}\label{thm:main}%
	Let everything be as in Lemma \ref{lem:main}. 
	
	The time $t$ price of a $T_0 \times (T_N-T_0)$ payer ($\zeta=1$) and receiver ($\zeta=-1$) swaption is given by
	\begin{align*}
		\Swaption_{T_0}^{T_N}(t;K,\zeta)&=
		P(t,T_0)\left(
			C_1 \normalCDF\left(\frac{C_1}{\sqrt{C_2}}\right)+
			\sqrt{C_2}\normalPDF\left(\frac{C_1}{\sqrt{C_2}}\right)
			\left(
				1 + \sum_{l=3}^{\infty}{(-1)^l q_l H_{l-2}}
			\right)
		\right),
	\end{align*}
	where $\mathcal{N}$ denotes the cdf of the normal distribution, $\normalPDF$ is the pdf or the normal distribution and $H_l$ are the probabilist's Hermite polynomials (see \Cref{sec:HermitePoly}). The coefficients
	$q_0=1$, $q_1=q_2=0$, and for $n\geq 3$
	\begin{align*}
		q_n =
		\sum_{m=1}^{\left\lfloor \frac{n}{3}\right\rfloor}{
			\sum_{\substack{k_1+\dots+k_m=n\\k_i\geq 3}}^{}{
				\frac{C_{k_1}\cdots C_{k_m}}{m!k_1!\cdots k_m!}
				\left(\frac{1}{\sqrt{C_2}}\right)^n
			}
		}
	\end{align*}
	for $C_l\coloneqq c_l(t) P(t,T_n)^l$ with $c_l(t)$ being the swap cumulants
	from Lemma \ref{lem:main} for fixed $t\geq 0$.
\end{theorem}
This formula will provide the necessary ingredient for the numerical experiments in \Cref{sec:numerics} making it possible to calibrate the model to the swaption surface very efficiently. After successfully calibrating the model, we apply it to find constant maturity swap rates in \Cref{sec:CMS} and Bermudan swaption pricing in \Cref{sec:bSwaption} using the Least-Square Monte Carlo technique. We will see a good performance of this model compared to the reference data downloaded from Bloomberg.
%%%%%%%%%%%%%%%%%%%%%%%%%%%%%%%%%%%%%%%%%%%%%%%%%%%%%%%%%%%%%%%%%%%%%%%%%%%%%%%%%%%%%%%%%%%
%% Review of the literature and comparison
%%%%%%%%%%%%%%%%%%%%%%%%%%%%%%%%%%%%%%%%%%%%%%%%%%%%%%%%%%%%%%%%%%%%%%%%%%%%%%%%%%%%%%%%%%%
\subsection{Review of the literature and comparison}\label{sec:review}
Historically, the theory of interest-rate modeling started on the assumption of specific one-dimensional dynamics for the instantaneous spot rate process $r$. 
These models are convenient for defining all fundamental quantities (rates and bonds) by no-arbitrage arguments as the expectation of a functional of the process $r$. 
Indeed, the price at time $t>0$ of a contingent claim with payoff $H_T$, $T>t$, under the risk-neutral measure $Q$ is given by (cf. \cite{PascucciBook})
\begin{align}\label{eq:riskneutralPrice}
H_t= E_t^{\Q} \left[e^{-\int_{t}^{T}r(s)ds} H_T\right],
\end{align}
where $E_t^\Q$ denotes the conditional expectation with respect to some filtration $\mathcal{F}_t$ under measure $\Q$. In particular, choosing $H_T\coloneqq P(T,T)=1$, where $P(t,T)$ denotes a zero-coupon bond.

The literature on interest rate modeling is very vast and our short literature review is by no means exhaustive. We refer to \cite{Bjork}, \cite{BrigoMercurioLibro} and \cite{HullLibro} for a comprehensive review and description of these models.  

Among all possible classifications, we can divide these models into two major categories: the endogenous and exogenous models. In chronological order, the first short-rate models belong to the first group: the Vasicek model \cite{Vasicek}, the Dothan model and the Cox, Ingersoll \& Ross (CIR) \cite{CIR}. In particular, the CIR model has been regarded as the reference model in interest rate modeling by both practitioners and academics for several decades for several reasons. First of all, it was derived from a general equilibrium framework. Secondly, it generates more realistic interest rate distributions with skewness and fatter tail with respect to normal distribution. Thirdly, it avoids negative interest rates. There is a rich literature on extensions to the classical CIR model in order to obtain more sophisticated models, which could fit the market data better, allowing to price interest rate derivatives more accurately. For example, Chen in \cite{Chen} proposed a three-factor model; Brigo and Mercurio in  \cite{BrigoMercurioLibro} proposed a jump diffusion model (JCIR). 

But in the last decade the financial industry encountered a paradigm shift by allowing
the possibility of negative interest rates, making the classical CIR model unsuitable.

Recently, Orlando et al. suggest in several papers (cf. \cite{Orlando2}, \cite{Orlando1} and  \cite{Orlando3}) a new framework, which they call CIR\# model, that fits the market term-structure of interest rates. Additionally, it preserves the market volatility, as well as the analytical tractability of the original CIR model. 
Their new methodology consists in partitioning the entire available market data sample, which usually consists of a mixture of probability distributions of the same type.
They use a technique to detect suitable sub-samples with normal or gamma distributions. In a next step, they calibrate the CIR parameters to shifted market interest rates, such that the interest rates are positive, and use a Monte Carlo scheme to simulate the expected value of interest rates.

Beside historical reasons, endogenous models are important for their simplicity and analytical tractability, in particular for the possibility of pricing bonds and bond options analytically. But there are some drawbacks. Since these models use only a few constant parameters, they are not able to reproduce simultaneously a given term-structure and volatility curve satisfactorily. Moreover, some shapes of the zero-coupon curve can never be
reproduced (for example an inverted shape curve with the Vasicek model). The need for an exact fit to the currently observed yield curve led some authors to introduce exogenous term-structure models. The first model was proposed by Ho \& Lee (see \cite{HoLee1986}), but we believe the most popular among practitioners is the Hull \& White extended Vasicek model (see \cite{HullWhite}).  A generalization of this model with a good calibration to swaption market prices was found in \cite{DiFrancesco}, while Mercurio and Pallavicini in \cite{MercurioPallavicini} proposed a mixing Gaussian model coupled with parameter uncertainty.

On the one hand, these models can handle negative interest rates with a very good analytical tractability. On the other hand, the distribution of continuously compounded interest rates shows all the undesirable features of the Gaussian distribution.

%%%%%%%%%%%%%%%%%%%%%%%%%%%%%%%%%%%%%%%%%%%%%%%%%%%%%%%%%%%%%%%%%%%%%%%%%%%%%%%%%%%%%%%%%%%
%% Contribution
In this paper, we extend the endogenous model of \cite{DiFrancesco2021} to an exogenous model by adding a deterministic shift and show how the Gram-Charlier expansion of \cite{Tanaka2010} can be utilized to calibrate our model to the swaption surface. We will see a good performance of the model with respect to determining constant maturity swap rates and pricing Bermudan swaptions.

We performed tests on two different dates \dateA and \dateB. At the first date, the market zero rates were partially negative and at the second date they were completely negative. We saw similar numerical results at both dates and decided for the sake of brevity to only present the results at \dateA. For the interested reader we will make the data at \dateB as well as the code of the numerical implementation available online.\footnote{For the numerical implementation and the data used in this paper please visit \url{https://github.com/kevinkamm/CIR--}.}
%%%%%%%%%%%%%%%%%%%%%%%%%%%%%%%%%%%%%%%%%%%%%%%%%%%%%%%%%%%%%%%%%%%%%%%%%%%%%%%%%%%%%%%%%%%
%% Structure of the paper
The paper is organized as follows. In \Cref{sec:model} we first introduce the deterministic shift extension and the corresponding zero-coupon price.
This is followed by a reminder on the relevant features of swaps and swaptions in \Cref{sec:swaptionPriceFormula}.

In \Cref{sec:GramCharlier} we will derive the Gram-Charlier expansion. This is done by first recalling how a probability density of a random variable can be approximated by Hermite polynomials. We will see that it is necessary to study the cumulants or equivalently the moments of this random variable. In our case, this will be the \emph{swap moments} and we will show, how to derive them from the so-called \emph{bond moments} by solving some Riccati equations, which will have explicit solutions in our model, making it possible to compute swaption prices very fast.

After that, in \Cref{sec:numerics}, we will conduct some numerical experiments. First, we calibrate our model to the market swaption surface at \dateA in \Cref{sec:calibration}. Subsequently, we simulate the model by using the Euler-Maruyama scheme in \Cref{sec:emc} and compute CMS rates in \Cref{sec:CMS}. We conclude our numerical tests by pricing Bermudan swaptions in \Cref{sec:bSwaption}.
Finally, we summarize the results of the paper in \Cref{sec:conclusion} and discuss possible extensions for future research.

%%%%%%%%%%%%%%%%%%%%%%%%%%%%%%%%%%%%%%%%%%%%%%%%%%%%%%%%%%%%%%%%%%%%%%%%%%%%%%%%%%%%%%%%%%%
%% The model
%%%%%%%%%%%%%%%%%%%%%%%%%%%%%%%%%%%%%%%%%%%%%%%%%%%%%%%%%%%%%%%%%%%%%%%%%%%%%%%%%%%%%%%%%%%
\section{A model for negative interest rates with perfect fit to the term-structure}\label{sec:model}
Let us define $\alpha\coloneqq \left(\alpha_x,\alpha_y\right)$, 
$\alpha_z\coloneqq \left(k_z,\theta_z,\sigma_z\right)$, $z\in \left\{x,y\right\}$. We want to use the general deterministic shift extension by 
\cite[pp.~95\,ff. Chapter 3.8 A General Deterministic-Shift Extension]{BrigoMercurioLibro} or \cite{Brigo2001} in the case of multifactor models. We note that contrary to the presented ideas in the aforementioned papers, we do not need to introduce another probability space for our purposes and will use the same risk-neutral measure for all dynamics.

Thus, we are interested in the following short rate model on $\left(\Omega,\mathcal{F},\Q\right)$
\begin{align}
	r(t;\alpha)\coloneqq \rCIR(t;\alpha) + \psi\left(t;\alpha\right) =
	x(t;\alpha_x)-y(t;\alpha_y) + \psi\left(t;\alpha_x\right), 
	\quad \psi\left(0;\alpha\right)=0,
	\label{eq:modelShiftExtended}
\end{align}
where $\rCIR$ denotes the short-rate model without the deterministic shift extension. We will suppress the dependency on the parameters $\alpha$ for readability whenever there is no confusion.

Likewise, we recall from Theorem \ref{thm:zcpriceCIR-} in the Appendix that the price of the zero-coupon bond for the non-extended model is given by
\begin{align*}
	\PCIR(t,T)=A_x(t,T)e^{-B_x(t,T)x(t)}A_y(t,T)e^{B_y(t,T)y(t)}.
\end{align*}
Analogue to \cite[p.~5 Theorem 3.1]{Brigo2001} we derive easily the price of a zero-coupon
bond in the short-rate model \eqref{eq:modelShiftExtended} for given parameters $\alpha$ 
\begin{align*}
	P(t,T)=
	\mathbb{E}^{\Q}\left[
		\left.
			\exp\left(
				-\int_{t}^{T}{
					r(s)ds
				}
			\right)
		\right|
		\mathcal{F}_t
	\right]&=
	\mathbb{E}^{\Q}\left[
		\left.
			\exp\left(
				-\int_{t}^{T}{
					x(s)-y(s)+\psi(s)ds
				}
			\right)
		\right|
		\mathcal{F}_t
	\right]\\&=
	\exp\left(
		-\int_{t}^{T}{
			\psi(s) ds
		}
	\right)
	\PCIR(t,T)
\end{align*}
because $\psi$ is deterministic.

To ensure a perfect fit to the initial term-structure, we set as in \cite[pp.~5--6 Corollary 3.2]{Brigo2001} 
\begin{align*}
	\psi(t;\alpha)=f^M(0,t)-f^\alpha(0,t),
\end{align*}
where $f^M(0,t)$ is the instantaneous market forward rate and 
\begin{align*}
f^\alpha(0,t)=
		-\frac{\partial_T\left(A_x(0,t)\right)}{A_x(0,t)}
		+\partial_T\left(B_x(0,t)\right) x(0)
		-\frac{\partial_T\left(A_y(0,t)\right)}{A_y(0,t)}
		-\partial_T\left(B_y(0,t)\right) y(0)
\end{align*}
is instantaneous market rate for $\rCIR$ (see \Cref{sec:instantForwardRate}).

More conveniently, we observe that this is equivalent to asking that the following equation holds
\begin{align*}
	\exp\left(
		-\int_{t}^{T}{
			\psi^\alpha(s) ds
		}
	\right)=
	\frac{P^M(0,T)}{P^M(0,t)}\frac{P^\text{CIR-}(0,t)}{P^\text{CIR-}(0,T)},
\end{align*}
where we used $P^z(t,T)=\exp\left(-\int_{t}^{T}{f^z(t,s)ds}\right)$,
$z\in \left\{M,\text{CIR-}\right\}$.

In total, this leads to the following formula for the zero-coupon price of the deterministic shift extended model
\begin{align}
	P(t,T)=
	\frac{P^M(0,T)}{P^M(0,t)}\frac{P^\text{CIR-}(0,t)}{P^\text{CIR-}(0,T)}
	P^\text{CIR-}(t,T)
	\label{eq:zcPrice--}
\end{align}
and $P(0,T)=P^M(0,T)$ is guaranteed.
%%%%%%%%%%%%%%%%%%%%%%%%%%%%%%%%%%%%%%%%%%%%%%%%%%%%%%%%%%%%%%%%%%%%%%%%%%%%%%%%%%%%%%%%%%%
%% Swaption Price Formula
%%%%%%%%%%%%%%%%%%%%%%%%%%%%%%%%%%%%%%%%%%%%%%%%%%%%%%%%%%%%%%%%%%%%%%%%%%%%%%%%%%%%%%%%%%%
\subsection{Swaption Price Formula}\label{sec:swaptionPriceFormula}
In \cite{DiFrancesco2021} we calibrated the short-rate model $\rCIR$ to the initial term-structure. For the deterministic shift extended model this is not possible, since a perfect fit is guaranteed regardless of the parameters $\alpha$. Therefore, we will calibrate to market swaption prices, for which we will recall all essentials in this section.

We are following 
\cite[\ pp.~428\,ff. Chapter 27.7 Swaps]{Bjork},
\cite[\ pp.~19\,ff.]{BrigoMercurioLibro} and 
\cite[\ pp.~3\,ff.]{Schrager2006} in this section.

A swap is a financial contract between two counterparties with fixed resettlement dates
$T_0,T_{1},\dots,T_N$, $N\in\N$. The contract itself contains two cashflows, one---called the floating leg---are payments of future interest rates and the other---called the fixed leg---is a fixed amount of payments. The receiver of a swap will receive at the fixed dates the amount of the fixed leg and pays the amounts of the floating leg to the other counterparty, giving it its name: the floating leg is swapped for the fixed leg.
Additionally, a payer swap refers to the case, when the floating leg is received and the fixed rate payed. We will distinguish the different kinds by introducing the factor 
$\swapType$, which will be equal to $+1$ in case of a payer swap and $-1$ in case of a receiver swap.

Such a contract with maturity $T_0$ and tenor
$T_N-T_0$ and resettlements $T_{0},\dots,T_N$  is commonly called a
$T_0 \times (T_N - T_0)$ swap.

The net value of a $T_0 \times (T_N - T_0)$ payer and receiver swap at time $t<T_0$ is given by
\begin{align}
	%\begin{aligned}[c]\arraycolsep=0pt
	\Swap^{T_N}_{T_0}(t;K,\swapType)&\coloneqq
	\swapType\left(P(t,T_0)-P(t,T_N)-K\sum_{i=1}^{N}{\alpha_i P(t,T_i)}\right)
	\label{eq:swapValue}
\end{align}
where $\alpha_i=T_{i}-T_{i-1}$ is the day-count convention and $K$ the fixed rate, see for instance \cite[\ pp.~429\,ff.]{Bjork}. To ease notation, we will suppress the explicit dependency on the $T_0$ and $T_N$ whenever there is no confusion.

A particular fixed rate $K$ called \emph{par or forward swap rate} is of special interest, which are usually quoted in the market.
It is the one, such that $\Swap^N(t;K,\swapType)=0$ (which is independent of $\swapType$) and we will denote it a bit more generally by
\begin{align*}
	\parSwapRate^N_n(t) \coloneqq
	\frac{P(t,T_n)-P(t,T_N)}{\sum_{i=n+1}^{N}{\alpha_i P(t,T_i)}},\quad n=0,\dots,N-1.
\end{align*}
Moreover, we will denote the so-called \emph{accrual factor} or \emph{present value of a basis point} by
\begin{align*}
	\PVBP^N_n(t)\coloneqq \sum_{i=n+1}^{N}{\alpha_i P(t,T_i)},\quad n=0,\dots,N-1.
\end{align*}

Now, we are able to discuss swaptions (cf. \cite[\ pp.~430\,ff.]{Bjork}). A 
$T_0 \times (T_N - T_0)$ payer, receiver swaption with swaption strike $K$ is a contract, which
at maturity $T_0$ gives the holder the right to enter into a 
$T_0 \times (T_N - T_0)$ payer, receiver swap with fixed rate $K$.

Its arbitrage free price at time $t<T_0$ is given by
\begin{align}
	\Swaption^{T_N}_{T_0}\left(t;K,\swapType\right)&=
	\mathbb{E}^{\Q}\left[
		\left.
		\exp\left(-\int_{t}^{T_0}{r(s)ds}\right)
		\left(\swapType\left(\parSwapRate^N_0(T_0)-K\right)\right)^+
		\PVBP^N_0(T_0)
		\right|
		\mathcal{F}_t
	\right]
	\label{eq:normalSwaption}
\end{align}
We will use this formulation for our Monte-Carlo calibration procedure together with
\eqref{eq:zcPrice--}. 
%%%%%%%%%%%%%%%%%%%%%%%%%%%%%%%%%%%%%%%%%%%%%%%%%%%%%%%%%%%%%%%%%%%%%%%%%%%%%%%%%%%%%%%%%%%
%% Dynamics under the swap measure
%%%%%%%%%%%%%%%%%%%%%%%%%%%%%%%%%%%%%%%%%%%%%%%%%%%%%%%%%%%%%%%%%%%%%%%%%%%%%%%%%%%%%%%%%%%
\subsubsection{Swaption prices under the forward measure}\label{sec:forwardMeasure}
For the Gram-Charlier expansion we will make use of the fact that the stochastic discount factor in \eqref{eq:normalSwaption} can be removed by a clever change of measure.
For fixed $T_0$, the $T_0$-forward measure $\Q^{T_0}$ is defined as the martingale measure for the numeraire process $p(t,T_0)$ (cf. \cite[pp.~403\,ff. Chapter 26.4 Forward measure]{Bjork}) and we have the following:

\begin{proposition}\label{prop:Björk268}%
	For any $T_0$-claim $X$ we have
	\begin{align*}
		\mathbb{E}^{\Q}\left[\left.e^{-\int_{t}^{T_0}{r(s)ds}}X\right|\mathcal{F}_t\right]=
		P(t,T_0)\mathbb{E}^{\Q^{T_0}}\left[\left.X\right|\mathcal{F}_t\right].
	\end{align*}
\end{proposition} 
Thus, the price at time $t<T_0$ of a payer ($\swapType=1$), receiver ($\swapType=-1$) swaption under the $T_0$-forward measure is given by
\begin{align}
	\Swaption\left(t;K,\swapType\right)&=
	P(t,T_0)
	\mathbb{E}^{\Q^{T_0}}\left[
		\left.
		\left(\swapType\left(\parSwapRate^N_0(T_0)-K\right)\right)^+
		\PVBP^N_0(T_0)
		\right|
		\mathcal{F}_t
	\right]
	%\begin{aligned}[c]\arraycolsep=0pt
	%\payerSwaption\left(t;K\right)&=
	%P(t,T_0)
	%\mathbb{E}^{\Q^{T_0}}\left[
		%\left.
		%\left(\parSwapRate^N_0(T_0)-K\right)^+
		%\PVBP^N_0(T_0)
		%\right|
		%\mathcal{F}_t
	%\right],\\
	%\receiverSwaption\left(t;K\right)&=
	%P(t,T_0)
	%\mathbb{E}^{\Q^{T_0}}\left[
		%\left.
		%\left(K-\parSwapRate^N_0(T_0)\right)^+
		%\PVBP^N_0(T_0)
		%\right|
		%\mathcal{F}_t
	%\right].
	%\end{aligned}
	\label{eq:forwardSwaption}
\end{align}
\section{Gram-Charlier expansion}\label{sec:GramCharlier}
We will use all the results available in \cite[pp.~3\,ff. Section 2.1 Gram-Charlier expansion]{Tanaka2010} and apply them to our case.

Let us first of all make the following observation: The payer ($\swapType=1$) and receiver ($\swapType=-1$) swap value
\eqref{eq:swapValue} can both be rewritten as
\begin{align*}
	\Swap(t;K,\swapType)\coloneqq
	\sum_{i=0}^{N}{
		a^\swapType_i P(t,T_i)
	},
\end{align*}
where $a^\swapType_i$ are equal to 
\begin{align*}
	a_0^\swapType&\coloneqq\swapType,& a_N^\swapType&\coloneqq-\swapType\left(1+K\alpha_{N}\right),& a_i^\swapType&\coloneqq-\swapType K\alpha_i,\quad i=1,\dots,N-1.
\end{align*}
For the remainder of this section we will drop the dependency on $\swapType$ for the coefficients $a_i$ to ease the notation.
Now, with this notation, we can rewrite the swaption prices \eqref{eq:forwardSwaption} to get
\begin{align*}
	\Swaption(t;K,\swapType)&=
	P(t,T_0)
	\mathbb{E}^{\Q^{T_0}}\left[
		\left.
		\left(\Swap^N(T_0;K,\swapType)\right)^+
		\right|
		\mathcal{F}_t
	\right]
	\\&\overset{!}{=}
	P(t,T_0)
	\int_{0}^{\infty}{
		x f(x) dx
	},
\end{align*}
for an unknown density function $f$. The idea of the Gram-Charlier expansion is to approximate this density function $f$ by using the orthonormal basis of Hermite polynomials (see Appendix \ref{sec:HermitePoly}), which is the content of the next Proposition (cf. \cite[p.~3 Proposition 2.1]{Tanaka2010} and \cite[p.~5 Proposition 2.1.2]{Cheng2013}).
\begin{proposition}\label{prop:GramCharlier}%
	Assume that a random variable $Y$ has the continuous density function $f$ and has
	finite cumulants $c_k$, $k\geq 1$. Then the following holds:
	\begin{compactenum}
		\item $f$ can be expanded as
			\begin{align*}
				f(x)=\sum_{n=0}^{\infty}{
					\frac{q_n}{\sqrt{c_2}} H_n\left(\frac{x-c_1}{\sqrt{c_2}}\right)
					\normalPDF\left(\frac{x-c_1}{\sqrt{c_2}}\right)
				},
			\end{align*}
			where $H_n$ are the probabilist's Hermite polynomials
			and $\normalPDF$ the probability density function of the standard normal distribution, as well as
			$q_0=1$, $q_1=q_2=0$, and for $n\geq 3$
			\begin{align*}
				q_n = 
				\frac{1}{n!} 
				\mathbb{E}\left[
					H_n\left(\frac{Y-c_1}{\sqrt{c_2}}\right)
				\right]=
				\sum_{m=1}^{\left\lfloor \frac{n}{3}\right\rfloor}{
					\sum_{\substack{k_1+\dots+k_m=n\\k_i\geq 3}}^{}{
						\frac{c_{k_1}\cdots c_{k_m}}{m!k_1!\cdots k_m!}
						\left(\frac{1}{\sqrt{c_2}}\right)^n
					}
				}.
			\end{align*}
		\item For any $a\in \R$
			\begin{multline*}
				\mathbb{E}\left[
					Y \1_{Y \geq a}
				\right]=
				c_1 \normalCDF\left(\frac{c_1-a}{\sqrt{c_2}}\right)+
				\sqrt{c_2} \normalPDF\left(\frac{c_1-a}{\sqrt{c_2}}\right)\\+
				\sum_{n=3}^{\infty}{
					\left(-1\right)^{n-1} q_n 
					\normalPDF\left(\frac{c_1-a}{\sqrt{c_2}}\right)
					\left[
						a H_{n-1}\left(\frac{c_1-a}{\sqrt{c_2}}\right)-
						\sqrt{c_2} H_{n-2}\left(\frac{c_1-a}{\sqrt{c_2}}\right)
					\right]
				},
			\end{multline*}
			where furthermore $\normalCDF$ denotes the cumulative distribution function of the standard normal distribution.
	\end{compactenum}
	In particular, we have
	\begin{align*}
		q_3=\frac{c_3}{3! c_2^{\frac{3}{2}}},&&
		q_4=\frac{c_4}{4! c_2^{\frac{5}{2}}},&&
		q_5=\frac{c_5}{5! c_2^{\frac{6}{2}}},&&
		q_6=\frac{c_6+10 c_3^2}{6! c_2^{\frac{6}{2}}},&&
		q_7=\frac{c_7+35c_3c_4}{7! c_2^{\frac{7}{2}}}
	\end{align*}
\end{proposition}
Therefore, all we have to do is determine the swap cumulants. This will be done in several steps: First, we will use the fact that cumulants can be computed from moments, see Appendix \ref{sec:cumulants}. Second, we compute the so-called swap moments in Equation \eqref{eq:swapMoments}, which in turn are computed from so-called bond moments. Last but not least, in order to compute the bond moments, we need to derive a new system of Riccati equations in Equation \eqref{eq:RiccatiTerminal}.
%%%%%%%%%%%%%%%%%%%%%%%%%%%%%%%%%%%%%%%%%%%%%%%%%%%%%%%%%%%%%%%%%%%%%%%%%%%%%%%%%%%%%%%%%%
%% Bond and Swap moments
%%%%%%%%%%%%%%%%%%%%%%%%%%%%%%%%%%%%%%%%%%%%%%%%%%%%%%%%%%%%%%%%%%%%%%%%%%%%%%%%%%%%%%%%%%
\subsection{Bond and Swap moments}\label{sec:moments}
Since, cumulants can be expressed by moments, vice versa (see Appendix \ref{sec:cumulants}), we will study the \emph{Swap moments} in this section, which we denote by
\begin{align*}
	M^m(t)\coloneqq
	\mathbb{E}^{\Q^{T_0}}\left[
		\left.
		\left(
			\Swap(T_0)
		\right)^m
		\right|
		\mathcal{F}_t
	\right]=
	\mathbb{E}^{\Q^{T_0}}\left[
		\left.
		\left(
		\sum_{i=0}^{N}{
			a_i P(T_0,T_i)
		}
		\right)^m
		\right|
		\mathcal{F}_t
	\right].
\end{align*}
It can be shown by induction that the $m$-th power can be rewritten as
\begin{align*}
	\left(
		\sum_{i=0}^{N}{
			a_i P(T_0,T_i)
		}
	\right)^m=
	\sum_{0\leq i_1,\dots,i_m\leq N}^{}{
		a_{i_1}\cdots a_{i_m}
		\left(
			\prod_{k=1}^{m}{
				P(T_0,T_{i_k})
			}
		\right)
	}.
\end{align*}
%a short proof via induction is given in Appendix \ref{sec:Misc}.

Now, notice that all $a_i$ are $\mathcal{F}_t$ measurable and therefore
\begin{align}
	M^m(t)=
	\sum_{0\leq i_1,\dots,i_m\leq N}^{}{
		a_{i_1}\cdots a_{i_m}
		\mathbb{E}^{\Q^{T_0}}\left[
			\left.
			\prod_{k=1}^{m}{
				P(T_0,T_{i_k})
			}
			\right|
			\mathcal{F}_t
		\right]
	}	
	\label{eq:swapMoments}
\end{align}
and we will call $\mathbb{E}^{\Q^{T_0}}\left[
			\left.
			\prod_{k=1}^{m}{
				P(T_0,T_{i_k})
			}
			\right|
			\mathcal{F}_t
		\right]$ the \emph{bond moments}.
		
Similar to \cite[\ pp.~44--46]{Cheng2013} we will reduce the problem to finding the bond moments for the short-rate model without a determinstic-shift extension by using 
\eqref{eq:zcPrice--}
\begin{align*}
	M^m(t)&=
	\sum_{0\leq i_1,\dots,i_m\leq N}^{}{
		a_{i_1}\cdots a_{i_m}
		\mathbb{E}^{\Q^{T_0}}\left[
			\left.
			\prod_{k=1}^{m}{
				P(T_0,T_{i_k})
			}
			\right|
			\mathcal{F}_t
		\right]
	}\\&=
	\sum_{0\leq i_1,\dots,i_m\leq N}^{}{
		a_{i_1}\cdots a_{i_m}
		\mathbb{E}^{\Q^{T_0}}\left[
			\left.
			\prod_{k=1}^{m}{
				\frac{P^M(0,T_{i_k})}{P^M(0,T_0)}\frac{P^\text{CIR-}(0,T_0)}{P^\text{CIR-}(0,T_{i_k})}
				P^\text{CIR-}(T_0,T_{i_k})
			}
			\right|
			\mathcal{F}_t
		\right]
	}\\&=
	\left(\frac{P^\text{CIR-}(0,T_0)}{P^M(0,T_0)}\right)^m
	\sum_{0\leq i_1,\dots,i_m\leq N}^{}{
		a_{i_1}\cdots a_{i_m}
		\prod_{k=1}^{m}{
			\frac{P^M(0,T_{i_k})}{P^\text{CIR-}(0,T_{i_k})}
		}
		\mathbb{E}^{\Q^{T_0}}\left[
			\left.
			\prod_{k=1}^{m}{
				P^\text{CIR-}(T_0,T_{i_k})
			}
			\right|
			\mathcal{F}_t
		\right]
	}\\&=
	\left(\frac{P^\text{CIR-}(0,T_0)}{P^M(0,T_0)}\right)^m
	\sum_{0\leq i_1,\dots,i_m\leq N}^{}{
		a^*_{i_1}\cdots a^*_{i_m}
		\mathbb{E}^{\Q^{T_0}}\left[
			\left.
			\prod_{k=1}^{m}{
				P^\text{CIR-}(T_0,T_{i_k})
			}
			\right|
			\mathcal{F}_t
		\right]
	},
\end{align*}
where $a^*_{i_k}=a_{i_k}\frac{P^M(0,T_{i_k})}{P^\text{CIR-}(0,T_{i_k})}$.

Thus, we only have to calculate the bond moments for the CIR- model.

For a numerical implementation, the $m$-fold sum over all permutations of $i_k$ is unfavorable.
Therefore, we rewrite it as follows: By definition, there will always be 
$m$ coefficients $a_{i_k}$ in the $m$-fold sum but it is possible to get e.g.
$a_{i_1}$ twice, etc. Hence, fixing indices for $a_0$ up to $a_N$ we can sum over the powers of all occurrences, which have to sum up to $m$. However, the individual products of the coefficients can appear multiple times as well, e.g. for $m=2, N=2$ summing over all permutations would lead to two times the term $a_0 a_1$, since we encounter $i_0=0,\ i_1=1$ and
$i_1=1,\ i_0=0$. 
Finally, we derive similar to \cite[\ p.~28 Remark 4.2.1]{Cheng2013} the following expression
\begin{align*}
	%\hspace{1em}&\hspace{-1em}
	\sum_{0\leq i_1,\dots,i_m\leq N}^{}{
		a_{i_1}\cdots a_{i_{m}} 
		\left(
			\prod_{k=1}^{m}{
				P(T_0,T_{i_k})
			}
		\right)
	}=
	\sum_{\substack{0\leq k_0,\dots,k_N\leq N\\k_0+\dots+k_N=m}}^{}{
		\frac{m!}{k_0!\cdots k_N!}
		a_{0}^{k_0}\cdots a_{N}^{k_N} 
		\left(
			\prod_{j=0}^{N}{
				P(T_0,T_{j})^{k_j}
			}
		\right)
	}.
\end{align*}

Finding this set of indices is known as \emph{subset sum problem}, which is NP-hard but can be solved by e.g. dynamical programming. The interested reader is referred to
\cite{Curtis2017} for recent developments using a GPU for large subset sum problems.
In our case, $m$ will be at most $7$ and due to annual payments $N$ will be at most equal to the maximal tenor plus one, i.e. $11$, which is considered as a small subset sum problem for which we will utilize a simpler implementation. Even with semi-annual payments a simple implementation with dynamic programming is sufficient, 
since we will need to calculate the subset sum problems only once and pass it to the calibration procedure.

Now, let us derive the Riccati equation for the bond moments. First of all, notice that the affine structure of $P(t,T)$ is preserved
\begin{align}
	\begin{aligned}[c]\arraycolsep=0pt
		\hspace{1em}&\hspace{-1em}
		\prod_{j=0}^{N}{
			\PCIR\left(T_0,T_{j}\right)^{k_j}
		}\\&=
		\prod_{j=0}^{N}{
			\left(
				A_x(T_0,T_{j}) e^{-B_x(T_0,T_{j}) x(T_0)}
				A_y(T_0,T_{j}) e^{B_y(T_0,T_{j}) y(T_0)}
			\right)^{k_j}
		}\\&=
		\left(\prod_{j=0}^{N}{
			A_x(T_0,T_{j})^{k_j} 
		}\right)
		e^{-\sum_{j=0}^{N}{k_j B_x(T_0,T_{j})}x(T_0)}
		\left(\prod_{j=0}^{N}{
			A_y(T_0,T_{j})^{k_j} 
		}\right)
		e^{\sum_{j=0}^{N}{k_j B_y(T_0,T_{j})}y(T_0)}
		\\&\eqqcolon
		A_x\left(T_0,\left\{k_{0},\dots,k_{N}\right\}\right)
		e^{-B_x\left(T_0,\left\{k_{0},\dots,k_{N}\right\}\right)x(T_0)}
		A_y\left(T_0,\left\{k_{0},\dots,k_{N}\right\}\right)
		e^{B_y\left(T_0,\left\{k_{0},\dots,k_{N}\right\}\right)y(T_0)}
	\end{aligned}
	\label{eq:bondMoments}
\end{align}
By Proposition \ref{prop:Björk268} we have also for $t\leq T_0$
\begin{align*}
	\hspace{1em}&\hspace{-1em}
	\mathbb{E}^{\Q^{T_0}}\left[
		\left.
		\prod_{j=0}^{N}{
			\PCIR\left(T_0,T_{j}\right)^{k_j}
		}
		\right|
		\mathcal{F}_t
	\right]\\&=
	\frac{1}{\PCIR\left(t,T_0\right)}
	\mathbb{E}^{\Q}\biggl[
		\biggl.
		e^{-\int_{t}^{T_0}{r^{\text{CIR-}}(s)ds}}
			A_x\left(T_0,\left\{k_{0},\dots,k_{N}\right\}\right)
		e^{-B_x\left(T_0,\left\{k_{0},\dots,k_{N}\right\}\right)x(T_0)}
		\\&\qquad
		A_y\left(T_0,\left\{k_{0},\dots,k_{N}\right\}\right)
		e^{B_y\left(T_0,\left\{k_{0},\dots,k_{N}\right\}\right)y(T_0)}
		\biggr|
			\mathcal{F}_t
	\biggr]\\
	&\overset{!}{=}
	\frac{1}{\PCIR\left(t,T_0\right)}
	M_x(t,T_0) e^{-N_x(t,T_0) x(t)}M_y(t,T_0) e^{N_y(t,T_0) y(t)}.
\end{align*}
We notice that by martingale pricing the discounted price process\\
$e^{-\int_{0}^{t}{\rCIR(s)ds}}
			M_x(t,T_0)
		e^{-N_x(t,T_0)}
		M_y(t,T_0)
		e^{N_y(t,T_0)}
$
has to be a martingale. Since it has an affine structure as well, it places us exactly in the same situation as in the derivation of Lemma \ref{lem:Riccati} seen in \cite{DiFrancesco2021} with the difference of variable terminal conditions. 

Therefore, we have the same Riccati equation but different terminal values dependent on $k_{0},\dots,k_{N}$.

For generic terminal values $a_z,b_z\in\R_{\geq 0}$ the explicit solution is given by
\begin{align}
	\begin{aligned}[c]\arraycolsep=0pt
	M_z(t,T_0)&=
		a_z
		\left(
			\frac{
				\phi_1^z \exp\left(\phi_2^z (T_0-t)\right)
			}{
				\phi_1^z+
				\phi_2^z \left(\exp\left(\phi_1^z (T_0-t)\right)-1\right)
					\left(1+b_z\left(\phi_1^z-\phi_2^z\right)\right)
			}
		\right)^{\phi_3^z},\quad &M_z(T_0,T_0)&=a_z\\
	N_z(t,T_0)&=
		\frac{
			b_z \phi_1^z + 
			\left(\exp\left(\phi_1^z (T_0-t)\right)-1\right)
				\left(1+b_z\left(\phi_1^z-\phi_2^z\right)\right)
		}{
			\phi_1^z+
				\phi_2^z \left(\exp\left(\phi_1^z (T_0-t)\right)-1\right)
					\left(1+b_z\left(\phi_1^z-\phi_2^z\right)\right)
		},\quad &N_z(T_0,T_0)&=b_z.
	\end{aligned}
	\label{eq:RiccatiTerminal}
\end{align}
As seen from our derivation in Equation \eqref{eq:bondMoments}, the terminal values $t=T_0$ are equal to
\begin{align*}
	a_z= A_z\left(T_0,\left\{k_{0},\dots,k_{N}\right\}\right),\quad
	b_z = B_z\left(T_0,\left\{k_{0},\dots,k_{N}\right\}\right), \quad z\in \left\{x,y\right\},
\end{align*}
and we can now compute the bond moments and therefore the swap moments for the Gram-Charlier expansion. Thus, using the one-to-one relationship between moments and cumulants in \Cref{sec:cumulants}, we have an explicit formula for the swap cumulants and we can apply
\Cref{prop:GramCharlier}, which is part of the next subsection.
%%%%%%%%%%%%%%%%%%%%%%%%%%%%%%%%%%%%%%%%%%%%%%%%%%%%%%%%%%%%%%%%%%%%%%%%%%%%%%%%%%%%%%%%%%%
%% Expansion formula
%%%%%%%%%%%%%%%%%%%%%%%%%%%%%%%%%%%%%%%%%%%%%%%%%%%%%%%%%%%%%%%%%%%%%%%%%%%%%%%%%%%%%%%%%%%
\subsection{Expansion formula}\label{sec:expansionFormula}
As described in \cite{Tanaka2010}, we can now use Proposition \ref{prop:GramCharlier} to formulate the Gram-Charlier expansion formula:
\begin{align*}
	\Swaption(t;K,\zeta)&=
	P(t,T_0)\left(
		C_1 \normalCDF\left(\frac{C_1}{\sqrt{C_2}}\right)+
		\sqrt{C_2}\normalPDF\left(\frac{C_1}{\sqrt{C_2}}\right)
		\left(
			1 + \sum_{l=3}^{\infty}{(-1)^l q_l H_{l-2}}
		\right)
	\right)
\end{align*}
where we replace the $c_n$ in Proposition \ref{prop:GramCharlier} by $C_n\coloneqq c_n(t) P(t,T_0)^n$ for $n\geq 1$ and the swap cumulants $c_n(t)$ are derived from the swap moments $M^m(t)$ using their one-to-one relationship shown in \Cref{sec:cumulants}.

In the following, we will denote by 
\begin{align*}
	\mathrm{GC}\left(L;K,\zeta\right)\coloneqq
	P(t,T_0)\left(
		C_1 \normalCDF\left(\frac{C_1}{\sqrt{C_2}}\right)+
		\sqrt{C_2}\normalPDF\left(\frac{C_1}{\sqrt{C_2}}\right)
		\left(
			1 + \sum_{l=3}^{L}{(-1)^l q_l H_{l-2}}
		\right)
	\right)
\end{align*}
the $L$-th order of the Gram-Charlier expansion of the $T_0 \times \left(T_N-T_0\right)$ swaption with strikes $K$ and swaption type $\zeta$ with annual payment dates .
%%%%%%%%%%%%%%%%%%%%%%%%%%%%%%%%%%%%%%%%%%%%%%%%%%%%%%%%%%%%%%%%%%%%%%%%%%%%%%%%%%%%%%%%%%%
%% Numerical Results
%%%%%%%%%%%%%%%%%%%%%%%%%%%%%%%%%%%%%%%%%%%%%%%%%%%%%%%%%%%%%%%%%%%%%%%%%%%%%%%%%%%%%%%%%%%
\section{Numerical tests}\label{sec:numerics}
%%%%%%%%%%%%%%%%%%%%%%%%%%%%%%%%%%%%%%%%%%%%%%%%%%%%%%%%%%%%%%%%%%%%%%%%%%%%%%%%%%%%%%%%%%%
%% Description of the Section
We will now perform some numerical experiments in our model.
In \Cref{sec:market_data} we will briefly discuss the market data, which we will use to perform all numerical tests in the subsequent sections. Afterwards, we will describe the calibration procedure of our model in \Cref{sec:calibration}. This is followed by a short subsection on simulating the model with the Euler-Maruyama scheme in \Cref{sec:emc} and in \Cref{sec:CMS} we investigate the par rates of constant maturity swaps (CMS). Last but not least, we compare the model Bermudan swaption prices to Bloomberg's Hull-White one factor model prices in \Cref{sec:bSwaption}.

%%%%%%%%%%%%%%%%%%%%%%%%%%%%%%%%%%%%%%%%%%%%%%%%%%%%%%%%%%%%%%%%%%%%%%%%%%%%%%%%%%%%%%%%%%%
%% Hardware and Software configuration
We used for the calculations \matlab with the \matlabGOtoolbox
running on \OS, on a machine with the following specifications: processor
\CPU and \RAM.
All calculations were sped-up by multiprocessing on a single CPU whenever possible.
%%%%%%%%%%%%%%%%%%%%%%%%%%%%%%%%%%%%%%%%%%%%%%%%%%%%%%%%%%%%%%%%%%%%%%%%%%%%%%%%%%%%%%%%%%%
%% Market Data
%%%%%%%%%%%%%%%%%%%%%%%%%%%%%%%%%%%%%%%%%%%%%%%%%%%%%%%%%%%%%%%%%%%%%%%%%%%%%%%%%%%%%%%%%%%
\subsection{Market Data}\label{sec:market_data}
To obtain the market zero-coupon bond term-structure, we first build the EUR Euribor-swap curve which is created from the most liquid interest rate instruments available in the market and constructed as follows: We consider deposit rates and Euribor rates with maturity from one day to one year and par-swap rates versus six-month Euribor rates with maturity from two years to thirty years. Then the zero interest curve and the zero-coupon bond curve are calculated using a standard ``bootstrapping'' technique in conjunction with cubic spline interpolation of the continuously compounded rate (cf. \cite{Miron1991} for more details). 

We tested the model at two different dates \dateA and \dateB. However, since the results on \dateA and \dateB are very similar, we decided to present only the results on \dateA and make all the data at \dateB available online to shorten the presentation.

We note that at \dateA (see \Cref{tab:first_curve}), the zero interest rates were negative up to year six, while at \dateB the entire zero interest rate structure was negative.

As aforementioned, we will calibrate the model to swaption prices (Table \ref{tab:swapPrices_1}).
They are computed by Bachelier's formula from normal volatilities quoted in the market 
(Table \ref{tab:swapVol_1}) and the swaption strikes can be found in \Cref{tab:swapStrike_1}.

After the calibration, we will assess the performance of the model by comparing its prediction of par CMS rates to Bloomberg's CMS rates in \Cref{sec:CMS}
and pricing Bermudan swaptions. 
The benchmark for Bermudan swaption prices will be Bloomberg's Hull-White one factor model alongside the corresponding strikes. The values are displayed in \Cref{tab:rbSwaptionHW_1}, \Cref{tab:pbSwaptionHW_1} and \Cref{tab:bSwaptionStrikes_1}, respectively.

All data has been downloaded from Bloomberg and is used in the following subsections for our numerical experiments. We start in the next subsection with calibrating our model to the swaption surface.
%%%%%%%%%%%%%%%%%%%%%%%%%%%%%%%%%%%%%%%%%%%%%%%%%%%%%%%%%%%%%%%%%%%%%%%%%%%%%%%%%%%%%%%%%%%
%% Calibration
%%%%%%%%%%%%%%%%%%%%%%%%%%%%%%%%%%%%%%%%%%%%%%%%%%%%%%%%%%%%%%%%%%%%%%%%%%%%%%%%%%%%%%%%%%%
\subsection{Calibration}\label{sec:calibration}
In this subsection we will discuss how we use the Gram-Charlier expansion to calibrate our model to parts of the swaption surface in Table \ref{tab:swapPrices_1}. Since we are using a deterministic shift extension, a perfect fit to the market zero-coupon curve (see Table \ref{tab:first_curve}) is always guaranteed.
%In this subsection we will discuss how we calibrate our model to the market zero-coupon curve given in \Cref{tab:first_curve} and \Cref{tab:second_curve} by using the formula derived in \eqref{eq:termstructure}.
%
Let us denote the parameter vector by $\Pi \coloneqq \left[\phi^x_1,\phi^x_2,\phi^x_3,\phi^y_1,\phi^y_2,\phi^y_3,x_0,y_0\right]^T \in \R^8_{>0}$. We will formulate the calibration procedure as a constraint minimization problem in $\R^8_{>0}$
for the parameters $\Pi$ with objective function
\begin{align}
	f(\Pi)\coloneqq
	\sum_{l\in \mathcal{L}}^{}{
		\sum_{T_0 \in \mathcal{M}}^{}{
			\sum_{T_N \in \mathcal{T}}^{}{
				\left(
					\frac{
						\mathrm{MarketSwaption}^{T_N}_{T_0}\left(K,\zeta\right)
					}{
						\mathrm{GC}^{T_N}_{T_0}\left(l,\Pi;K,\zeta\right)
					}-1
				\right)^2
			}
		}
	},
	\label{eq:objective}
\end{align}
where $\mathcal{L}\subset\N$ is a set of natural numbers containing the orders of the Gram-Charlier expansion, $\mathcal{M}$ is a set of maturities and $\mathcal{T}$ a set of final times. We will go into further details how to choose these sets in \Cref{rem:calibrationSets}.

The objective function describes the relative square difference between the market swaption prices and the theoretical prices derived by the Gram-Charlier expansion  using the short-rate model \eqref{eq:modelShiftExtended}.

The set of admissible parameters $\mathcal{A}$ will consist of the following constraints arising from the well-definedness of the formulas \eqref{eq:phi}:
\begin{compactenum}
	\item First of all, let us note that there is a one-to-one correspondence between the parameters $\Pi$ and $k_z$, $\sigma_z$ and $\theta_z$ if one is looking for positive real solutions only. We have
	\begin{align}
		\begin{aligned}[c]\arraycolsep=0pt
		k_x &= 2 \phi^x_2 - \phi^x_1,\qquad 
		&&k_y = 2 \phi^y_2 - \phi^y_1,\\
		\sigma_x &= \sqrt{2\left(\phi^x_2\phi^x_1-\left(\phi^x_2\right)^2\right)},\qquad
		&&\sigma_y = \sqrt{-2\left(\phi_2^y\phi_1^y-\left(\phi_2^y\right)^2\right)},\\
		\theta_x &= 
			-\frac{\phi^2_x\phi^3_x(\phi^1_x - \phi^2_x)}{\phi^1_x - 2\phi^2_x},\qquad
			%-(\phi^2_x*\phi^3_x*(\phi^1_x - \phi^2_x))/(\phi^1_x - 2*\phi^2_x)
		&&\theta_y = 
			\frac{\phi^2_y\phi^3_y(\phi^1_y - \phi^2_y)}{\phi^1_y - 2\phi^2_y}.
			%(\phi^2_y*\phi^3_y*(2*\phi^1_y - 2*\phi^2_y))/(2*(\phi^1_y - 2*\phi^2_y))
		\end{aligned}
		\label{eq:modelParam}
	\end{align}
	\item\label{item:cond1} We require  $\sigma_z \in \R_{\geq 0}$, $z\in \left\{x,y\right\}$. 
	 By rearranging \eqref{eq:modelParam}, these conditions are equivalent to
	$\phi_1^x \geq \phi_2^x$ and $\phi_2^y \geq \phi_1^y$;
	\item A positive mean-reversion speed, i.e. $k_z \geq 0$, is equivalent to 
		$2\phi_2^z \geq \phi_1^z$, $z\in \left\{a,b\right\}$;
	\item The Feller condition $2k_z\theta_z \geq \sigma_z^2 $ is equivalent to 
		$\phi_3^z\geq 1$, $z\in \left\{a,b\right\}$;
	\item A positive mean for each CIR process, i.e. $\theta_z \geq 0$, is 
		by positivity of $\sigma_z^2$ and $k_z$ equivalent to $\phi_3^{z}\geq 0$, which is already satisfied by the Feller condition;
	\item The parameter $\phi_1^z$, assuming that it is real-valued, is positive by definition, meaning that by the positivity of the mean reversion speed, $\phi_2^z$ will be as well. Therefore, all $\phi$ are positive;
	\item As both CIR processes $x_t$ and $y_t$, individually, are positive processes, we additionally require $x_0\geq 0$ and $y_0\geq 0$.
\end{compactenum}
The advantage of using the parameters $\Pi$ instead of $k_z$, $\sigma_z$ and $\theta_z$
is that we can rewrite these conditions as a system of linear inequality constraints in matrix notation $A\cdot\Pi\leq 0$, where
\begin{align*}
	A\coloneqq 
	\left[\begin{array}[c]{*{8}{c}}
    -1  &  1  &  0  &  0  &  0  &  0  &  0  &  0\\
     0  &  0  &  0  &  1  & -1  &  0  &  0  &  0\\
     1  & -2  &  0  &  0  &  0  &  0  &  0  &  0\\
     0  &  0  &  0  &  1  & -2  &  0  &  0  &  0\\
     %0  &  0  & -1  &  0  &  0  &  0  &  0  &  0\\
     %0  &  0  &  0  &  0  &  0  & -1  &  0  &  0\\
     %0  &  0  &  0  &  0  &  0  &  0  & -1  &  0\\
     %0  &  0  &  0  &  0  &  0  &  0  &  0  & -1		
	\end{array}\right]
\end{align*}
with boundary conditions $\Pi_i\geq 0$, $i=1,\dots,8$, and $\Pi_3=\phi_3^x\geq 1$, as well as $\Pi_6=\phi_3^y\geq 1$.

In total, the set of admissible parameters is given by 
\begin{align}
	\mathcal{A}\coloneqq 
	\left\{
		\Pi \in \R^8_{\geq 0}, \Pi_3, \Pi_6 \geq 1 : A \cdot \Pi \leq 0
	\right\}.
	\label{eq:constraints}
\end{align} 

Finally, a solution $\Pi^*$ to the calibration problem is a minimizer of
\begin{align}
	\min_{\Pi \in \mathcal{A}} f\left(\Pi\right).
	\label{eq:min}
\end{align}

Before we present some results, we would like to make the following remark on the choices of $\mathcal{L}$, $\mathcal{M}$ and $\mathcal{T}$.
\begin{remark}\label{rem:calibrationSets}%
	As always in calibration procedures with parametrized models, there is the notion of over- and underfitting to the data. Overfitting usually occurs when there are more parameters than independent values to calibrate to. For example, we saw a very good fit to a single swaption price. Underfitting on the other hand, occurs when the model is not able to fit to the whole data, e.g. fitting this model to the entire swaption surface.
	
	In our experiments, we determined that $4$ up to $6$ values performed best with regards to the Bermudan swaption pricing (\Cref{sec:bSwaption}) and finding the CMS par rates (\Cref{sec:CMS}).
	This is not very surprising, since the model has in total $8$ parameters but since the two CIR processes are independent and subtracted to deal with the negative interest rates it has essentially $4$ parameters to model the data.
	
	Therefore, we decided to perform tests on columns of the swaption surface and excluded short maturities. Additionally, removing the last maturity in the column from the calibration increased the speed of the optimization with usually the same accuracy.
	Additionally, we performed tests on several diagonals of the swaption surface with similar results and therefore decided to focus only on columns in this paper.
	
	Another aspect of this calibration procedure is the question which orders to use of the Gram-Charlier expansion. Since it is an orthogonal expansion, there is no a-priori error estimate of the truncated expansion formula. This also means that increasing the order might not be beneficial for the accuracy. Through comparing the Gram-Charlier swaption prices with Monte-Carlo swaption prices (see \Cref{tab:errMCGC}) using the same parameters, we found both prices to be closer too each other if we were using the order three, five and seven in the calibration procedure. A non-rigorous and heuristic idea behind this reasoning is that if the three orders are close too each other then the expansion ``converges'' to the correct price of the swaption in a loose sense.
	
	To conclude, to avoid over- and underfitting we will calibrate to columns of the swaption surface starting with maturity five and ending with maturity 15. Moreover, to have a ``stable'' Gram-Charlier swaption price we will use the orders three, five and seven in all experiments.
\end{remark}

To solve \eqref{eq:min} numerically, we would like to use \verb+Matlab+'s function \fmincon in
the \matlabGOtoolbox. In order to use this function, we need an initial guess of the parameter $\Pi$ and the computational time will depend on that choice.

Our experiments showed that initial guesses with small admissible values worked best for \fmincon. Therefore, we use the following hand-made parameters as initial points for \fmincon
\begin{align*}
	I_1 \coloneqq \left[0.1, 0.095, 0.3, 0.095, 0.1,0.3,0.01,0.01\right]^T, && I_2 \coloneqq \frac{1}{2} I_1
\end{align*}
and compare the performance to parameters found by \verb+Matlab+'s function \ga.
For the algorithms used by \verb+Matlab+ we refer the reader to \cite{Gilli}, in the context of financial mathematics.
 
In \Cref{tab:ctimeErrCal_1} we show the value of \eqref{eq:objective} after the calibration procedure and its computational time in seconds in the case of a payer swaption at \dateA. We display four different choices of initial points, first of all only using \ga, second \ga as an initial point for \fmincon, third $I_1$ as initial point for \fmincon and last but not least $I_2$ for \fmincon.
We can see that the model fits the swaption values best using columns with larger tenor but the computational time increases as well for all methods. Also we can see that our choices $I_1$ and $I_2$ in conjunction with \fmincon outperforms \ga with respect to accuracy and it is significantly faster than the combination of \ga and \fmincon. Therefore, we will use in the following experiments only \fmincon with $I_1$ or $I_2$ to present the results. In \Cref{tab:paramsFmincon1_1} we show the results of \eqref{eq:min} with initial point $I_1$ using \fmincon for reproducibility.
\begin{table}%
\centering
\caption{Computational times and values of \eqref{eq:objective} using different initial points and different swaption columns in \Cref{tab:swapPrices_1} and corresponding strikes \Cref{tab:swapStrike_1} in the case of payer swaptions and maturities ranging from 5 to 15.}
\begin{tabularx}{\linewidth}{@{}|c|*{5}{C}|@{}}
\hline
\diagbox{Method}{Tenor} & 1 & 2 & 5 & 7 & 10\\ 
\hline
\ga 							& \tabEC{3.94e-2}{76.2} & \tabEC{7.12e-2}{85.8} & \tabEC{5.75e-2}{100} 
										& \tabEC{2.27e-2}{168} & \tabEC{1.79e-2}{891}\\
\ga \& \fmincon 	& \tabEC{3.94e-2}{76.6} & \tabEC{7.92e-2}{87.6} & \tabEC{6.61e-3}{118.6} 
										& \tabEC{1.12e-3}{206.6} & \tabEC{8.04e-4}{945.9}\\
$I_1$ \& \fmincon & \tabEC{7.90e-2}{0.9} & \tabEC{4.78e-2}{0.8} & \tabEC{6.62e-3}{2.47} 
										& \tabEC{1.10e-3}{52} & \tabEC{3.00e-4}{181}\\
$I_2$ \& \fmincon & \tabEC{8.62e-1}{0.3} & \tabEC{5.80e-1}{1.35} & \tabEC{6.55e-3}{33.3} 
										& \tabEC{1.12e-3}{49.9} & \tabEC{6.95e-4}{93.9}\\
\hline
\end{tabularx}
\label{tab:ctimeErrCal_1}
\end{table}
\begin{table}%
\centering
\caption{Calibrated parameters $\Pi^*$ using $I_1$ with \fmincon and different swaption columns in \Cref{tab:swapPrices_1} and corresponding strikes \Cref{tab:swapStrike_1} in the case of payer swaptions and maturities ranging from 5 to 15.}
\begin{tabularx}{\linewidth}{@{}|c|*{5}{C}|@{}}
\hline
\diagbox{$\Pi^*$}{Tenor} & 1 & 2 & 5 & 7 & 10\\ 
\hline
$\phi_1^x$ & 0.082 		& 0.114  	& 0.109 	& 0.113 	 & 0.118\\ 
$\phi_2^x$ & 0.0477 	& 0.0947 	& 0.0846 	& 0.0899 	 & 0.092\\ 
$\phi_3^x$ & 1.05 		& 1.13 	 	& 1.99 		& 2 			 & 2\\ 
$\phi_1^y$ & 0.155 		& 0.0241 	& 0.584 	& 0.00192  & 0.00741\\ 
$\phi_2^y$ & 0.165 		& 0.0521 	& 0.597 	& 0.00851  & 0.00151\\ 
$\phi_3^y$ & 1.33 		& 1.19 	 	& 1.26 		& 1.78 		 & 1.73\\ 
$x_0$ 		 & 0.000126 & 0.00147 & 0.00017 & 0.000107 & 0.00151\\ 
$y_0$ 		 & 0.000128 & 0.0024 	& 0.0021 	& 0.0991 	 & 0.0988\\ 
\hline
\end{tabularx}
\label{tab:paramsFmincon1_1}
\end{table}

%%%%%%%%%%%%%%%%%%%%%%%%%%%%%%%%%%%%%%%%%%%%%%%%%%%%%%%%%%%%%%%%%%%%%%%%%%%%%%%%%%%%%%%%%%%
%% Monte Carlo Simulation
%%%%%%%%%%%%%%%%%%%%%%%%%%%%%%%%%%%%%%%%%%%%%%%%%%%%%%%%%%%%%%%%%%%%%%%%%%%%%%%%%%%%%%%%%%%
\subsection{Euler-Monte-Carlo simulation}\label{sec:emc} 
In order to forecast the future expected interest rate for e.g. pricing Bermudan swaptions in \Cref{sec:bSwaption}, we use the Euler-Maruyama scheme to simulate the instantaneous spot rate $r$ \eqref{eq:r}. We refer to \cite{Dereich} and the references therein for a list of different Euler-type methods to simulate a CIR process. In our experiments, we simulate the processes $x(t)$ and $y(t)$ by the truncated Euler scheme defined as follows:

First of all, we fix a homogeneous time grid $0=t_0\leq t_1 \leq \cdots \leq t_N=T$ 
for the interval $[0,T]$ with $N+1$ time points and mesh $\Delta t_i \coloneqq t_{i+1}-t_{i} \equiv \Delta \coloneqq \frac{T}{N}$ for all $i=0,\dots,N-1$. Secondly, we simulate the two independent Brownian motions $W_z$, $z\in \left\{x,y\right\}$, and define their time increment as $\Delta W_{z}(t_i)\coloneqq W_{z}(t_{i+1})-W_{z}(t_{i})$.
In total, we compute $r(t_{i+1})\coloneqq x(t_{i+1})-y(t_{i+1})$ for $i=0,\dots,N-1$, where
\begin{align}\label{eq:euler}
\begin{aligned}[c]\arraycolsep=0pt
 	x({t_{i+1}}) &= x({t_i}) + k_x(\theta_x - x({t_i}))\Delta t_i + \sigma_x\sqrt{\max(x({t_i}),0)}\Delta W_x({t_i}) \\
 	y({t_{i+1}}) &= y({t_i}) + k_y(\theta_y - y({t_i}))\Delta t_i + \sigma_y\sqrt{\max(y({t_i}),0)}\Delta W_y({t_i}).
\end{aligned}
\end{align}
We choose the $\max$ inside the square-root to ensure that the square-root remains real, because due to discretization effects the positivity of $x({t_i})$ and $y({t_i})$ might be violated.

In all of our experiments, we will use $M=10000$ simulations and mesh size $\Delta = \frac{1}{256}$.
On the one hand, looking at the fast calibration times using the Gram-Charlier approximation in \Cref{sec:calibration}, it is clear that Monte-Carlo methods cannot compete with respect to speed. On the other hand, since the Gram-Charlier expansion has no a-priori error bound let us now validate the calibration results by computing the Monte-Carlo prices with the parameters obtained by the Gram-Charlier expansion in \Cref{tab:paramsFmincon1_1}.
In \Cref{tab:errMCGC} we compare the swaption prices obtained by selected orders of the Gram-Charlier expansion to the Monte-Carlo prices and also the Monte-Carlo prices to the market prices. To compare the prices, we will use an average absolute error, i.e. for $X,Y \in \R^{d_1,d_2}$
\begin{align*}
	\norm{X-Y}\coloneqq 
	\frac{1}{d_1 d_2} \sum_{i=1}^{d_1}{\sum_{j=1}^{d_2}{\abs{X_{ij}-Y_{ij}}}}.
\end{align*}
\begin{table}%
\centering
\caption{Average absolute errors of Monte-Carlo prices compared to Gram-Charlier prices and market prices using the parameters shown in \Cref{tab:paramsFmincon1_1}.}
\begin{tabularx}{\linewidth}{@{}|c|*{5}{C}|@{}}
\hline
\diagbox{Methods}{Tenor} & 1 & 2 & 5 & 7 & 10\\ 
\hline
MC $-$ GC$3$ 	& 4.96e-4 & 1.17e-3 & 5.48e-4 & 6.23e-4 & 6.18e-4\\ 
MC $-$ GC$5$ 	& 2.90e-4 & 1.07e-4 & 9.54e-4 & 2.93e-4 & 3.57e-4\\ 
MC $-$ GC$7$ 	& 7.65e-4 & 1.05e-3 & 2.14e-4 & 1.97e-4 & 2.86e-4\\ 
MC $-$ Market & 3.93e-4 & 8.99e-4 & 4.58e-4 & 3.85e-4 & 3.73e-4\\ 
\hline
\end{tabularx}
\label{tab:errMCGC}
\end{table}
The average absolute error between the Gram-Charlier orders and the Monte-Carlo prices are usually of order $10^{-4}$ and the Monte-Carlo prices compared to the market prices usually of order $10^{-4}$, as well. It is important to note while reading this table that the prices themselves are usually of order $10^{-2}$, therefore the accuracy is usually up to two significant orders. Hence, this validates the parameters obtained by the calibration with the Gram-Charlier expansion and we can proceed with finding CMS rates in the next subsection using Monte-Carlo techniques.
%%%%%%%%%%%%%%%%%%%%%%%%%%%%%%%%%%%%%%%%%%%%%%%%%%%%%%%%%%%%%%%%%%%%%%%%%%%%%%%%%%%%%%%%%%%
%% Pricing CMS
%%%%%%%%%%%%%%%%%%%%%%%%%%%%%%%%%%%%%%%%%%%%%%%%%%%%%%%%%%%%%%%%%%%%%%%%%%%%%%%%%%%%%%%%%%%
\subsection{Pricing Constant Maturity Swaps (CMS)}\label{sec:CMS}
In this section, we want to use the calibrated model to compute the par rates of constant maturity swaps (CMS) using Monte-Carlo simulation. We refer the reader to \cite[pp.~557\,ff. Section 13.7 Constant-Maturity-Swaps]{BrigoMercurioLibro} and \cite[pp.~7\,ff.]{Tanaka2010} for more details.

Let us recall the definition of a CMS:
\begin{definition}\label{def:CMS}%
	A \emph{constant maturity swap (CMS)} is a variant of an interest rate swap between two parties, such that at each payment date starting at $T_0$ and ending at $T_N$ a fixed rate $K$ is swapped with a $c$-year swap rate.
	
	Analogously, we distinguish between payer and receiver CMS. In receiver CMS the fixed rate is received and the floating rate paid, vice versa for payer CMS.
\end{definition}
Also, as before, we will assume annual settlements between the effective date $T_0$ and maturity $T_N$ and denote the payment dates by $\mathcal{T}\coloneqq \left\{T_0,T_1,\dots,T_N\right\}$.
The net value of a $T_0 \times T_N + c$ CMS with fixed rate $K$ and index $c$ at time 0 under the risk-neutral measure is
\begin{align}
	\CMS^{T_N}_{T_0}\left(0;K,c,\zeta\right)\coloneqq
	\mathbb{E}^{\Q}\left[
		\sum_{i=1}^{N}{
			\exp\left(
				-\int_{0}^{T_{i-1}}{
					r(s) ds
				}
			\right)
			\zeta
			\alpha_i
			\left(
				R_{i-1}^{i-1+c}(T_{i-1}) - K
			\right)
		}
	\right].
	\label{eq:CMSvalue}
\end{align}
By rearranging \eqref{eq:CMSvalue}, we can compute the par CMS rates by setting it to zero and solve for $K$, i.e.
\begin{align*}
	K&=
	\frac{
		\mathbb{E}^{\Q}\left[
			\sum_{i=1}^{N}{
				\alpha_i
				\exp\left(
					-\int_{0}^{T_{i-1}}{
						r(s) ds
					}
				\right)
				R_{i-1}^{i-1+c}(T_{i-1})
			}
		\right]
	}{
		\mathbb{E}^{\Q}\left[
			\sum_{i=1}^{N}{
				\alpha_i
				\exp\left(
					-\int_{0}^{T_{i-1}}{
						r(s) ds
					}
				\right)
			}
		\right]
	}\\&=
	\frac{
		\mathbb{E}^{\Q}\left[
			\sum_{i=1}^{N}{
				\alpha_i
				\exp\left(
					-\int_{0}^{T_{i-1}}{
						r(s) ds
					}
				\right)
				R_{i-1}^{i-1+c}(T_{i-1})
			}
		\right]
	}{
			\sum_{i=1}^{N}{
				\alpha_i
				P(0,T_{i-1})
			}
	}.
\end{align*}
Remember that by the deterministic shift extension we have $P(0,T)=P^M(0,T)$ in our model.

In our experiment, we will use Monte-Carlo simulation for the short-rate \eqref{eq:r} and display the results in \Cref{tab:CMS_1} using the initial parameters $I_2$ for \fmincon in the case of payer swaptions. 
In the first column we see the effective date $T_0$, in the second the tenor $T$, such that $T_N=T_0+T$ and in the third column the index $c$ for the CMS. The next column shows Bloomberg's CMS rates, which is followed by the model CMS rates. In the last column we can see the absolute error of market and model rates. We can observe that the majority of CMS rates are very close to each other, telling us that the model performs well on average using just one column of the swaption data for the calibration. Using different columns in the calibration for all different CMS rates would improve the results further.
\begin{table}%
\centering
\caption{CMS rates computed with a calibration using $I_2$ and \fmincon to the column with tenor 7 of the payer swaption surface with maturities ranging from 5 to 15.}
\begin{tabular}{|*{6}{c}|}
\hline
Effective Date & Tenor & Index & Bloomberg's CMS Rate & Model CMS Rate & Abs Error\\
\hline
0 & 5 & 5 & 0.00145 & 0.00154 & 8.91e-05\\
0 & 10 & 5 & 0.00472 & 0.00499 & 0.000273\\
0 & 5 & 10 & 0.00465 & 0.0047 & 4.67e-05\\
0 & 10 & 10 & 0.00732 & 0.00738 & 6.06e-05\\
3 & 5 & 5 & 0.00562 & 0.00584 & 0.000226\\
3 & 5 & 10 & 0.00824 & 0.00825 & 8.32e-06\\
5 & 10 & 5 & 0.00999 & 0.01 & 3.64e-05\\
5 & 5 & 5 & 0.00958 & 0.00847 & 0.00112\\
5 & 5 & 10 & 0.011 & 0.0101 & 0.000912\\\hline
\end{tabular}
\label{tab:CMS_1}
\end{table}
%%%%%%%%%%%%%%%%%%%%%%%%%%%%%%%%%%%%%%%%%%%%%%%%%%%%%%%%%%%%%%%%%%%%%%%%%%%%%%%%%%%%%%%%%%%
%% Pricing Bermudan swaptions
%%%%%%%%%%%%%%%%%%%%%%%%%%%%%%%%%%%%%%%%%%%%%%%%%%%%%%%%%%%%%%%%%%%%%%%%%%%%%%%%%%%%%%%%%%%
\subsection{Pricing Bermudan swaptions}\label{sec:bSwaption}
In this section we want to use the calibrated model to compute the prices of Bermudan swaptions. A popular choice of literature on this subject is e.g \cite[pp.~588\,ff. Section 13.15 LFM: Pricing Bermudan Swaptions]{BrigoMercurioLibro}, \cite[pp.~421\,ff. Chapter 8 Pricing American Options]{Glasserman2004} or more recently \cite{Gatarek2021} and \cite[pp.~422\,ff. Section 13.3.2 European and Bermudan option example]{Oosterlee2019}.

Now, let us define which type of Bermudan swaptions we are interested in.
\begin{definition}\label{def:ncBermudanSwaption}%
	A \emph{$T_N$ no-call $T_0$} or 
		\emph{\abrBSwaption{T_N}{T_0} Bermudan swaption}
	with annual exercise dates
	gives its holder the right but not the obligation to enter at any time
	$\mathcal{T}_E^N \coloneqq \left\{T_0,T_{1},\dots,T_{N-1}\right\}$ into an
	interest rate swap with first reset $T\in \mathcal{T}_E$, last payment $T_N$
	and fixed rate $K$.
\end{definition}

Let us give a quick example of a 
\abrBSwaption{10}{2} Bermudan swaption with annually spaced exercise dates.
The holder can exercise this option starting from year 2 and afterwards at the beginning of each consecutive year but not later than year 9. After exercising the option, the holder enters into a swap contract---for simplicity with annual settlements---ending at year 10.

Accordingly, the price at time $t$ of a \abrBSwaption{T_N}{T_0} Bermudan swaption is the solution to the following optimal stopping problem
\begin{align*}
	\BSwaption^{T_N}_{T_0}\left(t;K,\swapType\right)\coloneqq
	\sup_{\substack{\tau \in \mathcal{T}_E^{N}\\
				\tau\text{ stopping time}}}
	\mathbb{E}^{\Q}_t\left[
		e^{-\int_{t}^{\tau}{r(s) ds}}
		S^N_\tau(\tau)
		\left(\swapType\left(K-R^N_\tau\left(\tau\right)\right)\right)^+
	\right],
\end{align*}
where the filtration is generated by the forward swap rate, i.e. 
$\mathcal{F}_t \coloneqq \sigma \left(R_s^N(s) : s\leq t\right)$ augmented such that it satisfies the usual hypothesis. 

For the implementation we are interested in the special case of today's price, i.e. $t=0$.
We will use backward induction to compare the exercise value to the continuation value and compute the conditional expectations by the least square Monte Carlo (LSMC) method (cf. \cite{Longstaff2001}).
Let us be more precise: 

%% Version without intermediate points:
We know that the price at time $T_{N-1}$ is given by 
\begin{align*}
	\BSwaption^{T_N}_{T_0}\left(T_{N-1};K,\swapType\right)&=
	\mathbb{E}^{\Q}_{T_{N-1}}\left[
		e^{-\int_{T_{N-1}}^{T_{N-1}}{r(s) ds}}
		S^N_{N-1}(T_{N-1})
		\left(\swapType\left(K-R^N_{N-1}\left(T_{N-1}\right)\right)\right)^+
	\right]\\&=
	P\left(T_{N-1},T_N\right)
	\left(\swapType\left(K-R^N_{N-1}\left(T_{N-1}\right)\right)\right)^+
\end{align*}
by definition and measurablility as well as the fact that the stopping time can only be equal to $T_{N-1}$ in this case. This gives us the opportunity to inductively calculate the Bermudan swaption price backwards.
 %First, let is introduce a time grid for $t$, i.e. $\mathcal{T}\coloneqq \left\{T_i \in [0,T_{N-1}] : T_i = i \frac{T_{N-1}}{I}, i=0,\dots I\right\}$, such that $\mathcal{T}^{n,N}_E \subseteq \mathcal{T}$. 
Thus, let us now assume that 
$\BSwaption^{T_N}_{T_0}\left(T_{i+1};K,\swapType\right)$ for $i=N-2,\dots,0$ is known.

We would like to compare the so-called \emph{continuation value}, which is the expected future payoff if the option is not exercised to the exercise value at all times $\mathcal{T}_E^{N}$, and is defined as
\begin{align*}
	c(T_i)\coloneqq
	\mathbb{E}^{\Q}_{T_{i}}\left[
		e^{-\int_{T_{i}}^{T_{i+1}}{r(s) ds}}
		\BSwaption^{T_N}_{T_0}\left(T_{i+1};K,\swapType\right)
	\right].
\end{align*}
Since the optimal stopping time will pathwise choose the maximum of continuing the option or exercising it, we have a dynamic programming principle
\begin{align*}
	\BSwaption^{T_N}_{T_0}\left(T_{i};K,\swapType\right)=
	\begin{cases}
		P\left(T_{N-1},T_N\right)
			\left(\swapType\left(K-R^N_{N-1}\left(T_{N-1}\right)\right)\right)^+, 
			& i=N-1\\ 
		\max\left(c(T_i),S^N_i(T_i)
			\left(\swapType\left(K-R^N_i\left(T_i\right)\right)\right)^+\right), 
			& i=0,\dots,N-2.\\ 
	\end{cases}
\end{align*}
The price at time $t=0$ is then given by
\begin{align*}
	\BSwaption^{T_N}_{T_0}\left(0;K,\swapType\right)=
	\mathbb{E}^{\Q}\left[
		e^{-\int_{0}^{T_{0}}{r(s) ds}}
		\BSwaption^{T_N}_{T_0}\left(T_{0};K,\swapType\right)
	\right].
\end{align*}

For completeness we explain how to approximate the conditional expectation with the LSMC method in Appendix \ref{sec:LSMC}. For the numerical implementation we choose the polynomial basis.

In \Cref{tab:avgErrBS_1} we can see the average absolute error of the Bermudan swaption prices in our model compared to Bloomberg's prices. We used as initial points $I_1$ and $I_2$ for \fmincon in the case of receiver and payer swaptions with different tenors. We can see that the average errors are very sensitive with respect to the calibrated parameters by looking at the results of $I_1$ and $I_2$ for a fixed tenor. Additionally, we notice that usually the results are better, if we choose $I_2$ as an initial point. The best results on average are found while calibrating to the columns of the swaption surface with tenor $5$ or $7$. In \Cref{tab:avgErrBS_1} we show the absolute errors for the individual payer Bermudan swaptions using $I_1$ as initial point calibrated to the column with tenor 5 and see an overall good match. Particularly, the column with tenor 7 in \Cref{tab:avgErrBS_1} is very accurate.

We focused in this experiment on the average errors only and not on specific Bermundan swaptions. If one desires to do so, there might be better choices which swaption prices to use for the calibration. Usually, the so-called co-terminal swaption prices are used to achieve better results for a specific Bermudan swaption. Since we are satisfied with the average performance of the model, we will not perform these individual tests for the sake of brevity.
\begin{table}%
\centering
\caption{Average absolute errors of Monte-Carlo Bermudan swaption prices and Bloomberg's HW1 Bermudan swaption prices using the $I_1$ and $I_2$ as initial points for \fmincon.}
\newsavebox{\diagMT}
\savebox{\diagMT}{\diagbox{Methods}{Tenor}}
\begin{tabularx}{\linewidth}{@{}|p{\wd\diagMT}|*{5}{C}|@{}}
\hline
\usebox{\diagMT} & 1 & 2 & 5 & 7 & 10\\ 
\hline
$I_1$ \& \fmincon \newline(Payer) 	 & 0.0254 	& 0.0129 	& 0.0014 	& 0.0073  & 0.203\\ 
$I_2$ \& \fmincon \newline(Payer) 	 & 0.00196 	& 0.00991 & 0.00269 & 0.00279 & 0.0088\\ 
$I_1$ \& \fmincon \newline(Receiver) & 0.948 		& 0.0642 	& 0.0036 	& 0.0021 & 0.0102\\ 
$I_2$ \& \fmincon \newline(Receiver) & 0.00615 	& 0.0149 	& 0.0033 	& 0.0021 & 0.0071\\ 
\hline
\end{tabularx}
\label{tab:avgErrBS_1}
\end{table}

\begin{table}%
\centering
\caption{Absolute errors of Monte-Carlo Bermudan payer swaption prices and Bloomberg's HW1 Bermudan swaption prices using the $I_1$ as initial points for \fmincon calibrated to the column with tenor equal to 5.}
\begin{tabular}{|c|*{4}{c}|}
\hline
\diagbox{Maturity}{Tenor} & 2 & 5 & 7 & 10\\
\hline
1 & 1.295e-03 & 5.452e-04 & 6.337e-04 & 2.488e-03\\
3 & 1.026e-03 & 6.628e-04 & 9.348e-04 & 2.931e-03\\
5 & 1.284e-03 & 1.404e-03 & 1.629e-04 & 3.605e-03\\
7 & 9.416e-04 & 1.191e-03 & 4.314e-06 & 2.271e-03\\
10 & 1.267e-03 & 1.305e-03 & 1.603e-03 & 2.470e-03\\\hline
\end{tabular}
\label{tab:absErrBS_1}
\end{table}
%%%%%%%%%%%%%%%%%%%%%%%%%%%%%%%%%%%%%%%%%%%%%%%%%%%%%%%%%%%%%%%%%%%%%%%%%%%%%%%%%%%%%%%%%%%
%% Conclusion and future research
%%%%%%%%%%%%%%%%%%%%%%%%%%%%%%%%%%%%%%%%%%%%%%%%%%%%%%%%%%%%%%%%%%%%%%%%%%%%%%%%%%%%%%%%%%%
\section{Conclusion and future research}\label{sec:conclusion}
In this paper, we extended the short-rate of \cite{DiFrancesco2021} by applying the deterministic-shift extension. We derived the swaption prices by using the Gram-Charlier expansion in this model and calibrated it to columns of the market swaption surface. The calibration is fast and accurate.
Using Monte-Carlo techniques, we obtained close CMS rates compared to Bloomberg's rates. Also compared to Bloomberg's Bermudan swaption prices via the HW1 model, our model performed very well. 

As a next step for future research, we would like to extend the model with piecewise constant coefficients and/or add more risk factors to capture more of the market swaption surface.
%%%%%%%%%%%%%%%%%%%%%%%%%%%%%%%%%%%%%%%%%%%%%%%%%%%%%%%%%%%%%%%%%%%%%%%%%%%%%%%%%%%%%%%%%%%
%% Appendix
%%%%%%%%%%%%%%%%%%%%%%%%%%%%%%%%%%%%%%%%%%%%%%%%%%%%%%%%%%%%%%%%%%%%%%%%%%%%%%%%%%%%%%%%%%%
\appendix
%%%%%%%%%%%%%%%%%%%%%%%%%%%%%%%%%%%%%%%%%%%%%%%%%%%%%%%%%%%%%%%%%%%%%%%%%%%%%%%%%%%%%%%%%%%
%%
%%%%%%%%%%%%%%%%%%%%%%%%%%%%%%%%%%%%%%%%%%%%%%%%%%%%%%%%%%%%%%%%%%%%%%%%%%%%%%%%%%%%%%%%%%%
\section{Results on the CIR- model}\label{sec:CIR-}
The following results are taken from \cite{DiFrancesco2021}.
\begin{theorem}\label{thm:zcpriceCIR-}%
	Let $\left(\Omega,\mathcal{F},\left(\mathcal{F}_t\right)_{t\in [0,T]},\Q\right)$ be a 
	stochastic basis, where $\Q$ is a martingale measure, $T>0$ a finite time horizon and let the $\sigma$-algebra $\left(\mathcal{F}_t\right)_{t\in [0,T]}$ fulfill the usual conditions and support two independent standard Brownian motions $W_x$ and $W_y$.
	
	The price of a zero-coupon bond in the model $r(t)=x(t)-y(t)$ 
	with $x$ and $y$ being two independent CIR processes as in \eqref{eq:CIR} is given by
	\begin{align}
	\PCIR(t,T) = A_x(t,T)e^{-B_x(t,T)x(t)}A_y(t,T)e^{B_y(t,T)y(t)},
		\label{eq:termstructure}
	\end{align}
	where $t\leq T$ and for $z\in \left\{x,y\right\}$
	\begin{align}\label{eq:AeB}
	\begin{aligned}[c]\arraycolsep=0pt
		A_z(t,T) &=  
			\left( 
				\frac{
					\phi^z_1 e^{\phi^z_2 (T-t)}
				}{
					\phi^z_2
						\left(
							e^{\phi^z_1(T-t)} -1
						\right)+ 
					\phi^z_1
				} 
			\right)^{\phi^z_3} \\
		B_z(t,T) &=  
			\frac{
				e^{\phi^z_1(T-t)}-1
			}{
				\phi^z_2 
					\left(
						e^{\phi^z_1(T-t)}-1
					\right)+ \phi^z_1
			}
	\end{aligned}
	\end{align} 
	with $\phi_i^z \geq 0$, 
	$i=1,2,3$, $z\in \left\{x,y\right\}$, such that the 
	Feller condition $2k_z\theta_z \geq \sigma_z^2$ is satisfied and
	\begin{align}\label{eq:phi}
	\begin{aligned}[c]%\arraycolsep=0pt
	&\phi^x_1 = \sqrt{k_x^2 + 2\sigma_x^2}, &&& 
	&\phi^x_2 = \frac{k_x + \phi^x_1}{2},&&&
	&\phi^x_3 = \frac{2k_x\theta_x}{\sigma_x^2} \\
	&\phi^y_1 = \sqrt{k_y^2 - 2\sigma_y^2},&&&
	&\phi^y_2 = \frac{k_y + \phi^y_1}{2}, &&&
	&\phi^y_3 = \frac{2k_y\theta_y}{\sigma_y^2}.
	\end{aligned}
	\end{align}
\end{theorem}
\begin{lemma}\label{lem:Riccati}%
	Let everything be as in Theorem \ref{thm:zcpriceCIR-} but let $x(t)$ and $y(t)$ follow the general
	affine dynamics 
	\begin{align}
	&\left\{
	\begin{aligned}[c]\arraycolsep=0pt
		dx(t)&=\left(\lambda_x(t) x(t) + \eta_x(t)\right)dt + 
			\sqrt{\gamma_x(t) x(t) + \delta_x(t)}dW_x(t)\\
		x(0)&=x_0,
	\end{aligned}
	\right.\label{eq:x}\\
	&\left\{
	\begin{aligned}[c]\arraycolsep=0pt
		dy(t)&=\left(\lambda_y(t) y(t) + \eta_y(t)\right)dt + 
			\sqrt{\gamma_y(t) y(t) + \delta_y(t)}dW_y(t)\\
		y(0)&=y_0,
	\end{aligned}
	\right.\label{eq:y}
\end{align}
The initial values $x_0,\ y_0\in \R$ are real-valued constants and the coefficients $\lambda_z,\eta_z,\gamma_z,\delta_z$, $z\in\left\{x,y\right\}$, are all real-valued deterministic functions, such that \eqref{eq:x} and \eqref{eq:y} are well-defined.
	
	Then, the price of a Zero-coupon bond is given by
	\begin{align}
		P(t,T)=
		E^\Q_t\left[e^{-\int_{t}^{T}r(s) ds}\right]=
		A_x(t,T)e^{-B_x(t,T)x(t)}A_y(t,T)e^{B_y(t,T)y(t)},
		\label{eq:PtTaffine}
	\end{align}
	where $A_z$ and $B_z$, $z\in \left\{x,y\right\}$, are deterministic functions and are a classical solution to the following system of Riccati equations
	\begin{align}
	\left\{
	\begin{aligned}[c]\arraycolsep=0pt
		-1 - B_x(t,T) \lambda_x(t) - \left(\partial_t B_x\right)(t,T) 
				+\frac{1}{2}B_x^2(t,T)\gamma_x(t)&=0,\quad B_x(T,T)=0\\
		 -B_x(t,T)\eta_x(t) + \frac{1}{2} B_x^2(t,T) \delta_x(t) + \partial_t\left(\log A_x\right)(t,T)&=0,\quad A_x(T,T)=1\\
		1 + B_y(t,T) \lambda_y(t) + \left(\partial_t B_y\right)(t,T) 
				+\frac{1}{2}B_y^2(t,T)\gamma_y(t)&=0,\quad B_y(T,T)=0\\
		 B_y(t,T)\eta_y(t) + \frac{1}{2} B_y^2(t,T) \delta_y(t) + \partial_t\left(\log A_y\right)(t,T)&=0,\quad A_y(T,T)=1.
	\end{aligned}
	\right.
	\label{eq:RiccatiEq}
	\end{align}
\end{lemma}
The Riccati equations for the CIR processes are given by
defining $\lambda_z(t)\equiv -k_z, \eta_z(t)\equiv k_z\theta_z, \gamma_z(t)\equiv\sigma_z^2, \delta_z(t)\equiv 0$

%Also notice that in this case
%\begin{align*}
	%\left(\partial_t A_x\right)(t,T) &= A_x(t,T) (k_x \theta_x) B_x(t,T),\\
	%\left(\partial_t A_y\right)(t,T) &= -A_y(t,T) (k_y \theta_y) B_y(t,T).
%\end{align*}
%%%%%%%%%%%%%%%%%%%%%%%%%%%%%%%%%%%%%%%%%%%%%%%%%%%%%%%%%%%%%%%%%%%%%%%%%%%%%%%%%%%%%%%%%%%
%% Instantaneous Forward Rate
%%%%%%%%%%%%%%%%%%%%%%%%%%%%%%%%%%%%%%%%%%%%%%%%%%%%%%%%%%%%%%%%%%%%%%%%%%%%%%%%%%%%%%%%%%%
\section{Instantaneous forward rate}\label{sec:instantForwardRate}
The definition of the instantaneous forward rate (cf. \cite{BrigoMercurioLibro} p.~13 equation (1.23)) is given by
\begin{align*}
	f(t,T)\coloneqq
	-\partial_T\log\left(P\left(t,T\right)\right).
\end{align*}
By \eqref{eq:termstructure} we therefore have
\begin{align*}
	f(t,T)&=
	-\partial_T\left(\log\left(A_x(t,T)e^{-B_x(t,T) x(t)}A_y(t,T)e^{B_y(t,T)y(t)}\right)\right)\\&=
	-\partial_T
		\left(
		\log\left(A_x(t,T)\right)-B_x(t,T) x(t)
		\right)
	-\partial_T
		\left(
		\log\left(A_y(t,T)\right)+B_y(t,T)y(t)
		\right)\\&=
		-\frac{\partial_T\left(A_x(t,T)\right)}{A_x(t,T)}
		+\partial_T\left(B_x(t,T)\right) x(t)
		-\frac{\partial_T\left(A_y(t,T)\right)}{A_y(t,T)}
		-\partial_T\left(B_y(t,T)\right) y(t).
\end{align*}
Let $z\in \left\{x,y\right\}$ and consider the case of the CIR model \eqref{eq:CIR}. Then those derivatives are given by the following expressions:
	Let us calculate the derivative of $A_z$ first
		\begin{align*}
			\hspace{1em}&\hspace{-1em}
			\partial_T\left(A_z(t,T)\right)\\&=
    \phi^3_z\left(\frac{\phi^1_z\phi^2_z{e}^{\phi^2_z\left(T-t\right)}}{\phi^1_z+\phi^2_z\left({e}^{\phi^1_z\left(T-t\right)}-1\right)}-\frac{\left(\phi^1_z\right)^2\phi^2_z{e}^{\phi^1_z\,\left(T-t\right)}{e}^{\phi^2_z\left(T-t\right)}}{{\left(\phi^1_z+\phi^2_z\left({e}^{\phi^1_z\left(T-t\right)}-1\right)\right)}^2}\right){\left(\frac{\phi^1_z{e}^{\phi^2_z\left(T-t\right)}}{\phi^1_z+\phi^2_z\left({e}^{\phi^1_z\left(T-t\right)}-1\right)}\right)}^{\phi^3_z-1}.
		\end{align*}
	Hence, we get
		\begin{align*}
			-\frac{\partial_T\left(A_z(t,T)\right)}{A_z(t,T)}=
			\frac{
				\phi^2_z \phi^3_z \left(\phi^1_z-\phi^2_z\right) \left({e}^{(T-t)\phi^1_z}-1\right)
			}{
				\phi^1_z+\phi^2_z\left({e}^{(T-t)\phi^1_z}-1\right)
			}.
		\end{align*}
	Now, we compute the derivative of $B_z$
		\begin{align*}
			\partial_T\left(B_z(t,T)\right)=
				\frac{
					\left(\phi^1_z\right)^2 {e}^{(T-t) \phi^1_z}
				}{
					{\left(\phi^1_z+\phi^2_z\left({e}^{(T-t) \phi^1_z}-1\right)\right)}^2
				}.
		\end{align*}
\section{Hermite Polynomials}\label{sec:HermitePoly}
In this short section we briefly recall the probabilist's Hermite polynomials, which are key to the Gram-Charlier expansion.
\begin{definition}\label{def:hermitePoly}%
	The \emph{(probabilist's) Hermite polynomials} $H_n(x)$ are defined as $H_0(x)\equiv 1$ and for 
	$n\geq 1$
	\begin{align*}
		(-1)^n \left(\normalPDF(x)\right)^{-1} \left(\frac{d^n}{dx^n} \normalPDF\right)(x),
	\end{align*}
	where $\normalPDF(x)\coloneqq \frac{1}{\sqrt{2\pi}}\exp\left(-\frac{x^2}{2}\right)$.
	
	Notice, that they are orthogonal with respect to the Gaussian measure, i.e.
	\begin{align*}
		\int_{\R}^{}{H_m(x)H_n(x)\normalPDF(x)dx}=\delta_{nm} n!.
	\end{align*}
	In particular,
	\begin{align*}
		&H_1(x)=x,
		H_2(x)=x^2-1,
		H_3(x)=x^3-3x,
		H_4(x)=x^4-6x^2+3,\\&
		H_5(x)=x^5-10x^3+15x,
		H_6(x)=x^6-15x^4+45x^2-15,
		H_7(x)=x^7-21x^5+105x^3-105x.
	\end{align*}
\end{definition}
%%%%%%%%%%%%%%%%%%%%%%%%%%%%%%%%%%%%%%%%%%%%%%%%%%%%%%%%%%%%%%%%%%%%%%%%%%%%%%%%%%%%%%%%%%%
%% Cumulants
%%%%%%%%%%%%%%%%%%%%%%%%%%%%%%%%%%%%%%%%%%%%%%%%%%%%%%%%%%%%%%%%%%%%%%%%%%%%%%%%%%%%%%%%%%%
\section{Cumulants and Moments}\label{sec:cumulants}
Let us denote by $\mu_i$ the moments and by $c_i$ the cumulants. Their relationship towards each other is determined by the moment generating function (cf. \cite{Smith1995}) like follows
\begin{align*}
	M(t)&=1+\sum_{i=1}^{\infty}{\mu_i \frac{t^i}{i!}}
			= \exp\left(\sum_{i=1}^{\infty}{c_i \frac{t^i}{i!}}\right)
			= \exp\left(K(t)\right).
\end{align*}
Therefore, assuming that the moments $\mu_i$ are known we can compute the cumulants $c_i$ by differentiating the formula from above 
\begin{align*}
	c_i = \left.\frac{d^i}{d t^i} \log\left(M(t)\right)\right|_{t=0}.
\end{align*}
Since, we only need a few of them, we can compute the formulas and implement them directly, leading to
\begin{align*}
	&c_1=\mu _1,\quad
	c_2=\mu _2-\mu _1^2,\quad
	c_3=2 \mu _1^3-3 \mu _2 \mu _1+\mu _3,\quad
	c_4=-6 \mu _1^4+12 \mu _2 \mu _1^2-4 \mu _3 \mu _1-3 \mu _2^2+\mu _4,
	\\&
	c_5=24 \mu _1^5-60 \mu _2 \mu _1^3+20 \mu _3 \mu _1^2+30 \mu _2^2 \mu _1-5 \mu _4 \mu _1-10 \mu _2 \mu _3+\mu _5\\&
	c_6=-120 \mu _1^6+360 \mu _2 \mu _1^4-120 \mu _3 \mu _1^3-270 \mu _2^2 \mu _1^2+30 \mu _4 \mu _1^2+120 \mu _2 \mu _3 \mu _1-6 \mu _5 \mu _1+30 \mu _2^3-10 \mu _3^2
	\\&\qquad -15 \mu _2 \mu _4+\mu _6\\&
	c_7=720 \mu _1^7-2520 \mu _2 \mu _1^5+840 \mu _3 \mu _1^4+2520 \mu _2^2 \mu _1^3-210 \mu _4 \mu _1^3-1260 \mu _2 \mu _3 \mu _1^2+42 \mu _5 \mu _1^2-630 \mu _2^3 \mu _1
	\\&\qquad+140 \mu _3^2 \mu _1+210 \mu _2 \mu _4 \mu _1-7 \mu _6 \mu _1+210 \mu _2^2 \mu _3-35 \mu _3 \mu _4-21 \mu _2 \mu _5+\mu _7.
\end{align*}

%%%%%%%%%%%%%%%%%%%%%%%%%%%%%%%%%%%%%%%%%%%%%%%%%%%%%%%%%%%%%%%%%%%%%%%%%%%%%%%%%%%%%%%%%%%
%% Least Square Monte Carlo method
%%%%%%%%%%%%%%%%%%%%%%%%%%%%%%%%%%%%%%%%%%%%%%%%%%%%%%%%%%%%%%%%%%%%%%%%%%%%%%%%%%%%%%%%%%%
\section{Least Square Monte Carlo method (LSMC)}\label{sec:LSMC}
In this section, we will demonstrate how to approximate the conditional expectation via LSMC. Let us first of all recall the following facts about the conditional expectation (cf. \cite[pp.~654\,ff.]{PascucciBook}):

Let $X\in L^2\left(\Omega,\mathcal{F},\Q\right)$ and $\mathcal{A}\subseteq \mathcal{F}$ be a 
		sub-$\sigma$-algebra. 
		
\begin{compactenum}
	\item Then the conditional expectation is the $L^2$-best approximation, i.e.
		\begin{align*}
			\mathbb{E}^{\Q}\left[
				\left(X-\mathbb{E}^{\Q}\left[\left.X\right|\mathcal{A}\right]\right)^2
			\right]\leq 
			\mathbb{E}^{\Q}\left[\left(X-Y\right)^2\right]
		\end{align*}
		for all $Y\in L^2\left(\Omega,\mathcal{A},\Q\right)$.
	\item Furthermore, the factorization Lemma tells us that there exists a function
		$u$, such that
		\begin{align*}
			\mathbb{E}^{\Q}\left[\left.Y\right|R\right]=u(R)
		\end{align*}
		and combined with the argument above
		\begin{align*}
			u(R)=\argmin_{v(\cdot)} \mathbb{E}^{\Q}\left[\abs{v(R)-Y}^2\right]
		\end{align*}
		where $v(\cdot)$ runs over all measurable functions.
\end{compactenum}

The idea is now to approximate the function $u(x)$. Therefore, fix a basis
$\left(b_i(x)\right)_{i=1,\dots,n}$ and 
set $b^n(\cdot)\coloneqq [b_1(\cdot),\dots,b_n(\cdot)]$.
Then, we approximate $u$ by $u(x)\approx \lambda^T b^n(x)$ where
$\lambda$ solves the least square problem
\begin{align*}
	\lambda = 
	\argmin_{\alpha\in \R^n} 
	\mathbb{E}^{\Q}\left[
		\abs{\alpha^Tb^n(R)-Y}^2
	\right].
\end{align*}

The problem we encounter is that in this least square problem we have random variables.
So we can numerically deal with this problem by simulating those random variables, if it is possible, and view this least square problem as finding a linear regression for data points introduced by the realizations of the random variables.

Thus, let $y_i$ be realizations of $Y$ and set $y=[y_1,\dots,y_m]^T$. Additionally, let 
$b_{ij}=b_i(r_j)$, where $r_j$ is a realization of $R$, and define the matrix 
$b=[b_{ij}]_{i=1,\dots,n;j=1,\dots,m}$.

Then the above least square problem reads
\begin{align*}
	\lambda = 
	\argmin_{\alpha\in \R^n} 
		\abs{b\alpha-y}^2.
\end{align*}
This is known as ordinary least square problem and the optimal solution
is given by
\begin{align*}
	\lambda=
	\left(b^Tb\right)^{-1}b^Ty.
\end{align*}
This tells us how to approximate the conditional expectation via a Monte Carlo linear regression approach.
%%%%%%%%%%%%%%%%%%%%%%%%%%%%%%%%%%%%%%%%%%%%%%%%%%%%%%%%%%%%%%%%%%%%%%%%%%%%%%%%%%%%%%%%%%%
%% Market data
%%%%%%%%%%%%%%%%%%%%%%%%%%%%%%%%%%%%%%%%%%%%%%%%%%%%%%%%%%%%%%%%%%%%%%%%%%%%%%%%%%%%%%%%%%%
\section{Market data}\label{sec:data}
\begin{table}[!ht]
	\caption{Market data containing the volatility surface for the swaption pricing at \dateA in bps.}
	\centering
	\small
	\volSwaptionA
	\label{tab:swapVol_1}
\end{table}

\begin{table}[!ht]
	\caption{Market data containing the swaption strikes at \dateA.}
	\centering
	\small
	\strikeSwaptionA
	\label{tab:swapStrike_1}
\end{table}

\begin{table}[!ht]
	\caption{Market data containing the swaption prices at \dateA.}
	\centering
	\small
	\marketSwaptionA
	\label{tab:swapPrices_1}
\end{table}

\begin{table}[!ht]
	\caption{Bloomberg's Hull-White one factor prices of receiver Bermudan swaptions at \dateA.}
	\centering
	\small
	\receiverbSwaptionHWA
	\label{tab:rbSwaptionHW_1}
\end{table}
\begin{table}[!ht]
	\caption{Bloomberg's Hull-White one factor prices of payer Bermudan swaptions at \dateA.}
	\centering
	\small
	\payerbSwaptionHWA
	\label{tab:pbSwaptionHW_1}
\end{table}
\begin{table}[!ht]
	\caption{Market data containing the Bermudan swaption strikes at \dateA.}
	\centering
	\small
	\bSwaptionStrikesA
	\label{tab:bSwaptionStrikes_1}
\end{table}
\begin{table}[!ht]
	\caption{Market data containing the zero rate curve and zero coupon curve at \dateA.}
	\centering
	\small
	\marketDataA
	\label{tab:first_curve}
\end{table}

%%%%%%%%%%%%%%%%%%%%%%%%%%%%%%%%%%%%%%%%%%%%%%%%%%%%%%%%%%%%%%%%%%%%%%%%%%%%%%%%%%%%%%%%%%%
%%
%%%%%%%%%%%%%%%%%%%%%%%%%%%%%%%%%%%%%%%%%%%%%%%%%%%%%%%%%%%%%%%%%%%%%%%%%%%%%%%%%%%%%%%%%%%
\section*{Declarations}
\subsection*{Funding}
This project has received funding from the European Union’s Horizon 2020 research and innovation
programme under the Marie Sklodowska-Curie grant agreement No 813261 and is part of the ABC-EU-XVA project.
\subsection*{Conflicts of interests}

The authors have no relevant financial or non-financial interests to disclose.

\subsection*{Data availability}
All data generated or analysed during this study are included in this published article.
{%\color{red}
In particular the code to produce the numerical experiments is available at\\
\url{https://github.com/kevinkamm/CIR--}.
}
%%%%%%%%%%%%%%%%%%%%%%%%%%%%%%%%%%%%%%%%%%%%%%%%%%%%%%%%%%%%%%%%%%%%%%%%%%%%%%%%%%%%%%%%%%%%
%% bibliography
%%%%%%%%%%%%%%%%%%%%%%%%%%%%%%%%%%%%%%%%%%%%%%%%%%%%%%%%%%%%%%%%%%%%%%%%%%%%%%%%%%%%%%%%%%%%
%{\thispagestyle{scrheadings}
%%\nocite{*}
%\newpage
%\thispagestyle{scrheadings}\ihead{}
%\printbibliography%[heading=bibintoc]
%}
{\thispagestyle{scrheadings}
%\nocite{*}
\newpage
\thispagestyle{scrheadings}\ihead{}
\singlespacing
\begin{footnotesize}
\bibliographystyle{acm}%{chicago}
\bibliography{literature.bib}
\end{footnotesize}
}

\end{document}